\documentclass[aps, prl]{revtex4-2}

\usepackage{amsmath}
\usepackage{circuitikz}
\usepackage{url}
\usepackage{geometry}
\begin{document}

\title{Johnson Noise Suppression in AC-Biased Transition-Edge Sensor Bolometers: A Thévenin-Equivalent Circuit Analysis}
\author{Tijmen de Haan}
\email{email: \url{tijmen.dehaan@gmail.com}}
\affiliation{Institute of Particle and Nuclear Studies (IPNS), High Energy Accelerator Research Organization (KEK), Tsukuba, Ibaraki 305-0801, Japan}
\affiliation{International Center for Quantum-field Measurement Systems for Studies of the Universe and Particles (QUP-WPI), High Energy Accelerator Research Organization (KEK), Tsukuba, Ibaraki 305-0801, Japan}
\date{August 29, 2025}

\begin{abstract}
\noindent
Transition-edge sensor (TES) bolometers operate under strong electrothermal feedback, wherein power deposited on the bolometer is compensated by a corresponding change in electrical power dissipation. We present a comprehensive analysis of Johnson noise suppression that extends previous theoretical frameworks to the general case of AC-biased linear circuits with arbitrary parasitic impedances. Using a Thévenin-equivalent circuit formulation—consisting of an ideal voltage source and complex series impedance external to the bolometer thermal island—we derive analytical expressions for the noise-equivalent current in the presence of both bolometer and parasitic Johnson noise sources. Our analysis demonstrates that while the electrothermal feedback loop effectively suppresses Johnson noise originating from the bolometer resistance, it provides no suppression of Johnson noise from external series resistance. In the limit of high loop gain, the bolometer Johnson noise contribution vanishes while the parasitic contribution remains unsuppressed.
\end{abstract}

\maketitle

\section{Introduction}

Transition-edge sensor (TES) bolometers have become the detector of choice for numerous applications (e.g. measurements of the cosmic microwave background) due to their exceptional sensitivity and ability to be manufactured in monolithic arrays \citep{mather_bolometer_1982}. These devices operate under strong voltage bias conditions \citep{lee_voltage-biased_1998, irwin_application_1995}, where negative electrothermal feedback (ETF) provides stability, linearity, speed-up of the time constant, and noise suppression. 

Johnson noise, arising from the thermal motion of charge carriers in resistive elements, represents a fundamental noise source in electrical circuits. In TES bolometers operated under strong voltage bias, the Johnson noise associated with the bolometer resistance experiences significant suppression through the electrothermal feedback mechanism \citep{de_wit_impact_2021}. As mentioned in \citet{irwin_transition-edge_2005}, the external electrical bias performs work against voltage fluctuations induced by random thermal motions, depositing heat that causes a corresponding change in the TES resistance. The voltage fluctuation and subsequent resistance fluctuation produce opposing currents that largely cancel at the readout amplifier, typically a superconducting quantum interference device (SQUID).

While previous analyses have considered Johnson noise suppression under idealized current or voltage bias conditions, practical readout systems inevitably contain parasitic impedances that modify these ideal bias conditions \citep{dobbs_frequency_2012}. Several mechanisms can contribute to these parasitic effects. First, normal-metal components in the signal path introduce series resistance—whether from cabling between the TES and its LC resonator multiplexer, or from inductive/capacitive coupling to nearby normal conductors. Furthermore, crosstalk between multiplexed TES channels creates mutual impedances, as each channel presents a non-zero impedance at other channels' bias frequencies. Yet another common cause of series impedance is from loss (ESR) in the multiplexing inductors or capacitors. Finally, and often most significantly in current-generation systems, common inductances downstream of the multiplexer summing point introduce effective series impedances that cannot be eliminated through bias frequency selection alone. These residual impedances fundamentally alter the electrothermal feedback dynamics and must be incorporated into any comprehensive noise analysis.

In this work, we extend the theoretical framework to encompass arbitrary AC bias configurations by employing a Thévenin-equivalent circuit representation. This approach provides a unified treatment applicable to all linear bias circuits without loss of generality.

\section{Theoretical Framework}
\subsection{Thévenin-Equivalent Circuit Representation}

According to Thévenin's theorem, any linear electrical circuit can be represented by an equivalent circuit consisting of an ideal AC voltage source, $V_{\mathrm{Th\acute{e}v}}$, in series with a complex impedance $z_{\mathrm{Th\acute{e}v}} = R_{\mathrm{Th\acute{e}v}} + iX_{\mathrm{Th\acute{e}v}}$. The resistive component of this impedance, $R_{\mathrm{Th\acute{e}v}}$, generates its own Johnson noise, characterized by a voltage noise source $V_{J,\mathrm{Th\acute{e}v}}$.

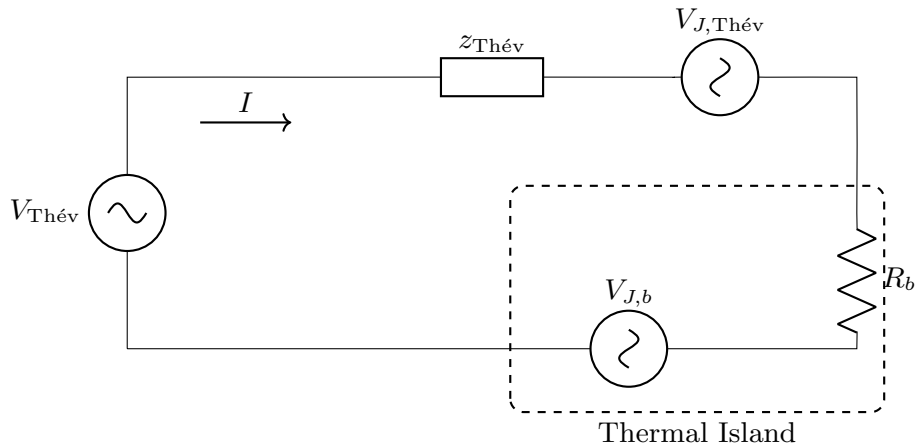
\begin{figure}[ht]
\centering
\begin{circuitikz}[scale=1.2, transform shape]
    % Define the circuit path
    \draw (0,0) 
        to[sV, l=$V_{\mathrm{Th\acute{e}v}}$] (0,3)
        to[short] (2,3)
        to[generic, l=$z_{\mathrm{Th\acute{e}v}}$] (6,3)
        to[short] (6,3)
        to[sV, l=$V_{J,\mathrm{Th\acute{e}v}}$] (7,3)
        to[short] (8,3)
        to[short] (8,1.5)
        to[R, l=$R_b$] (8,0)
        to[short] (6.5,0)
        to[sV, l_=$V_{J,b}$] (4.5,0)
        to[short] (0,0);
    
    % Add the thermal island box
    \draw[dashed, thick, rounded corners=5pt] (4.2,-0.7) rectangle (8.3,1.8);
    \node[anchor=north] at (6.25,-0.7) {\small Thermal Island};
    
    % Add current direction arrow
    \draw[->, thick] (0.8,2.5) -- (1.8,2.5);
    \node[above] at (1.3,2.5) {$I$};
\end{circuitikz}
\caption{Thévenin-equivalent circuit representation of a TES bolometer readout system. The circuit consists of a series loop containing: (1) an AC voltage source $V_{\mathrm{Th\acute{e}v}}$, (2) a complex Thévenin impedance $z_{\mathrm{Th\acute{e}v}} = R_{\mathrm{Th\acute{e}v}} + iX_{\mathrm{Th\acute{e}v}}$, (3) the Johnson noise source $V_{J, \mathrm{Th\acute{e}v}}$ associated with the resistive component $R_{\mathrm{Th\acute{e}v}}$, (4) the bolometer resistance $R_b$, and (5) the bolometer Johnson noise source $V_{J,b}$. The dashed boundary delineates the thermal island, which encompasses only $R_b$ and $V_{J,b}$.}
\label{fig:circuit}
\end{figure}

The bolometer has a temperature-dependent resistance $R_b$ with an associated Johnson noise voltage source $V_{J,b}$. Crucially, only these two elements reside on the thermal island and are therefore subject to the power balance equation that governs the electrothermal feedback mechanism. Figure~\ref{fig:circuit} illustrates the complete circuit topology: a single current loop containing the Thévenin voltage source, the Thévenin impedance with its noise source, and the bolometer with its corresponding noise source.

\subsection{Linearized Analysis}

We define the total circuit impedance as $Z = R_b + z_{\mathrm{Th\acute{e}v}}$ and perform a first-order perturbation analysis in the small-signal regime. The zeroth-order quantities comprise $V_{\mathrm{Th\acute{e}v}}$, $R_b$, and $z_{\mathrm{Th\acute{e}v}}$, while the first-order perturbations include the Johnson noise voltages $V_{J,b}$ and $V_{J, \mathrm{Th\acute{e}v}}$, and the induced resistance change $\delta R_b$. Higher-order terms are neglected throughout this analysis.

The current in the circuit can be expressed as $I = I_0 + \delta I$, where:
\begin{equation}
 I_0 = \frac{V_{\mathrm{Th\acute{e}v}}}{Z}
\end{equation}
represents the steady-state current, and the fluctuation is given by:
\begin{equation}
\delta I = \underbrace{\frac{V_{J,b}}{Z}}_{\substack{\text{Bolometer} \\ \text{Johnson}}} + \underbrace{\frac{V_{J,\mathrm{Th\acute{e}v}}}{Z}}_{\substack{\text{Parasitic} \\ \text{Johnson}}} - \underbrace{\frac{I_0 \cdot \delta R_b}{Z}}_{\substack{\text{ETF} \\ \text{Response}}}
\end{equation}

Notably, at this stage, the two Johnson noise sources appear symmetrically in the current fluctuation expression. The asymmetry emerges only when we consider the power balance on the thermal island.

\section{Power Balance and Electrothermal Feedback}

The electrical power dissipated on the thermal island induces a temperature change, which modulates the TES resistance and thereby alters the current flow—this constitutes the electrothermal feedback mechanism. For fluctuations much slower than the bolometer thermal time constant (the quasi-equilibrium regime), the power fluctuation $\delta P_b$ must be balanced by the change in heat flow $G \delta T$, where $G$ represents the dynamic thermal conductance.

The extension to finite-frequency effects is straightforward: the power balance equation must be modified to include a finite heat capacity. For the majority of the equations derived here, this is equivalent to modifying the loop gain $\mathcal{L}$ to $\mathcal{L}/(1+i\omega \tau)$, where $\tau$ is the thermal time constant. We present the low-frequency analysis here.

The power dissipation fluctuation on the thermal island is determined by the real part of the voltage-current product:
\begin{equation}
    \delta P_b = \Re \left( \delta \left( V_\mathrm{island}  I^* \right) \right) = G \delta T = \frac{G T \delta R_b}{\alpha R_b}
\end{equation}
where $\alpha = d\ln R/d\ln T$ is the logarithmic temperature coefficient of resistance, and the voltage across the island is:
\begin{equation}
    V_\mathrm{island} = V_\mathrm{Th\acute{e}v} + V_{J,\mathrm{Th\acute{e}v}} - z_\mathrm{Th\acute{e}v} (I_0 + \delta I)
\end{equation}

After retaining only first-order terms and noting that Johnson noise voltages $V_J$ are treated as phase-independent white noise amplitudes with $\Re (V_J I) = |I| V_{J}$, we obtain:
\begin{equation}
  \delta P_b = \frac{V_{\mathrm{Th\acute{e}v}}}{|Z|^2}  \left(R_b - R_{\mathrm{Th\acute{e}v}} \right) \left( V_{J,\mathrm{Th\acute{e}v}} + V_{J,b} \right)
  + \frac{(|z_{\mathrm{Th\acute{e}v}}|^2 - R_b^2) V_{\mathrm{Th\acute{e}v}}^2 \delta R_b}{|Z|^4}
  + \frac{V_{J,\mathrm{Th\acute{e}v}} V_{\mathrm{Th\acute{e}v}}}{|Z|} 
\end{equation}

Following \citet{de_haan_mntes_2024}, we introduce the electrothermal feedback loop gain:
\begin{equation}
\mathcal{L} = \frac{\alpha V_\mathrm{Th\acute{e}v}^2}{G T_b} \frac{R_b \left( R_b^2 - \left|  z_\mathrm{Th\acute{e}v} \right |^2 \right)}{|Z|^4 }
\end{equation}
where we have set $\beta=\partial R / \partial I=0$, which is a good approximation for many TES bolometers used for cosmic microwave background observations (e.g. \citet{zhou_method_2024}).

This yields the simplified first-order power balance equation:
\begin{equation}
\frac{V_{\mathrm{Th\acute{e}v}}}{|Z|^2} \left(R_b - R_{\mathrm{Th\acute{e}v}} \right) \left( V_{J,\mathrm{Th\acute{e}v}} + V_{J,b} \right) + \frac{V_{J,\mathrm{Th\acute{e}v}} V_{\mathrm{Th\acute{e}v}}}{|Z|} = \frac{G T}{\alpha R_b}(1 + \mathcal{L}) \delta R_b
\label{eq:power_balance}
\end{equation}

\section{Noise-Equivalent Current Analysis}

The total noise-equivalent current (NEI) power spectral density comprises contributions from both Johnson noise sources, which we treat as incoherent:
\begin{equation}
  \mathrm{NEI}^2 = 
  \left| \delta I \big|_{V_{J,b}\neq 0, V_{J,\mathrm{Th\acute{e}v}}=0} \right|^2 +
  \left| \delta I \big|_{V_{J,b}=0, V_{J,\mathrm{Th\acute{e}v}}\neq 0} \right|^2
\end{equation}
where $V_{J,b}^2 = 8kT_b R_b$ and $V_{J,\mathrm{Th\acute{e}v}}^2 = 8kT_{\mathrm{Th\acute{e}v}} R_{\mathrm{Th\acute{e}v}}$.

\subsection{Bolometer Johnson Noise Contribution}

For Johnson noise originating on the thermal island ($V_{J,b} \neq 0$, $V_{J,\mathrm{Th\acute{e}v}} = 0$), solving Equation~\ref{eq:power_balance} yields:
\begin{equation*}
\delta R_b = \frac{\mathcal{L}}{1+\mathcal{L}}\frac{V_{J,b}}{V_{\mathrm{Th\acute{e}v}}} \frac{R_b^2 \left( R_b-R_{\mathrm{Th\acute{e}v}} \right)}{R_b^2 - |z_{\mathrm{Th\acute{e}v}}|^2}
\end{equation*}

The resulting current fluctuation is:
\begin{equation}
\delta I_b = \frac{V_{J,b}}{Z} \left[1 - 
\frac{\mathcal{L}}{1+\mathcal{L}} \frac{R_b^2 \left( R_b-R_{\mathrm{Th\acute{e}v}} \right)}{Z(R_b^2 - |z_{\mathrm{Th\acute{e}v}}|^2)}
\right]
\end{equation}

In the limit of negligible parasitic impedance, this reduces to the familiar result $\delta I_b = V_{J,b}/(R_b(1+\mathcal{L}))$, demonstrating the suppression of bolometer Johnson noise by the factor $(1+\mathcal{L})$ under strong voltage bias.

\subsection{Parasitic Johnson Noise Contribution}

For Johnson noise from the Thévenin resistance ($V_{J,b} = 0$, $V_{J,\mathrm{Th\acute{e}v}} \neq 0$), the power balance equation yields:
\begin{equation*}
\delta R_b = \frac{\mathcal{L}}{1+\mathcal{L}}\frac{V_{J,\mathrm{Th\acute{e}v}}}{V_{\mathrm{Th\acute{e}v}}} \frac{R_b^2 \left( R_b-R_{\mathrm{Th\acute{e}v}} + |Z| \right)}{R_b^2 - |z_{\mathrm{Th\acute{e}v}}|^2}
\end{equation*}

The corresponding current fluctuation is:
\begin{equation}
\delta I_{\mathrm{Th\acute{e}v}} = \frac{V_{J,\mathrm{Th\acute{e}v}}}{Z} \left[1 - 
\frac{\mathcal{L}}{1+\mathcal{L}} \frac{R_b^2 \left( R_b-R_{\mathrm{Th\acute{e}v}} + |Z| \right)}{Z(R_b^2 - |z_{\mathrm{Th\acute{e}v}}|^2)}
\right]
\end{equation}

For small parasitic impedance, this yields $\delta I_{\mathrm{Th\acute{e}v}} = V_{J,\mathrm{Th\acute{e}v}}(1-\mathcal{L})/(R_b(1+\mathcal{L}))$. This demonstrates a key difference from the bolometer noise. When $\mathcal{L} = 1$, the in-phase component of parasitic signals is canceled by ETF (the quadrature component remains unchanged). At high loop gain, the original noise amplitude is restored, though with an (inconsequential) $180^\circ$ phase shift relative to the initial perturbation.

\section{Results and Discussion}

Combining the contributions from both Johnson noise sources, we obtain the total noise-equivalent current power spectral density:

\begin{equation}
\label{eq:result}
\boxed{
\begin{aligned}
\mathrm{NEI}^2 = &\frac{8kT_b R_b}{|Z|^2} \left|1 - \frac{\mathcal{L}}{1+\mathcal{L}} \frac{R_b^2 \left( R_b-R_{\mathrm{Th\acute{e}v}} \right)}{Z(R_b^2 - |z_{\mathrm{Th\acute{e}v}}|^2)} \right|^2 \\
&+ \frac{8kT_{\mathrm{Th\acute{e}v}} R_{\mathrm{Th\acute{e}v}}}{|Z|^2} \left|1 - \frac{\mathcal{L}}{1+\mathcal{L}} \frac{R_b^2 \left( R_b-R_{\mathrm{Th\acute{e}v}} + |Z| \right)}{Z(R_b^2 - |z_{\mathrm{Th\acute{e}v}}|^2)} \right|^2
\end{aligned}
}
\end{equation}

This expression is valid under the assumptions that Johnson noise voltages are small compared to the bias voltage and that the bolometer remains within its linear response regime. No assumptions have been made regarding the magnitude of the parasitic impedance.

For the practical case where the bias frequency is chosen to minimize reactance ($X_{\mathrm{Th\acute{e}v}} \approx 0$) and the parasitic resistance is small ($R_{\mathrm{Th\acute{e}v}} \ll R_b$), a Taylor expansion yields:
\begin{equation}
\mathrm{NEI}^2 = \frac{8kT_b}{R_b(1+\mathcal{L})^2} \left[ 1 + \frac{R_{\mathrm{Th\acute{e}v}}}{R_b} \left(4 \mathcal{L} - 2 + \frac{T_{\mathrm{Th\acute{e}v}}}{T_b}(1 - \mathcal{L})^2\right) + \mathcal{O}\left(\left(\frac{R_{\mathrm{Th\acute{e}v}}}{R_b}\right)^2\right) \right]
\end{equation}

Equation~\ref{eq:result} is the main result of this work and should be used for practical purposes. However, this expansion is insightful because it reveals the three first-order effects caused by the series impedance: (1) the modification of the detector responsivity, (2) the increased impedance to noise current, and (3) introduction of Johnson noise from the parasitic resistance itself. Crucially, in the limit of high loop gain ($\mathcal{L} \gg 1$), the bolometer Johnson noise remains strongly suppressed while the parasitic Johnson noise persists unsuppressed.

\begin{figure}[ht]
    \centering
    \includegraphics[width=\linewidth]{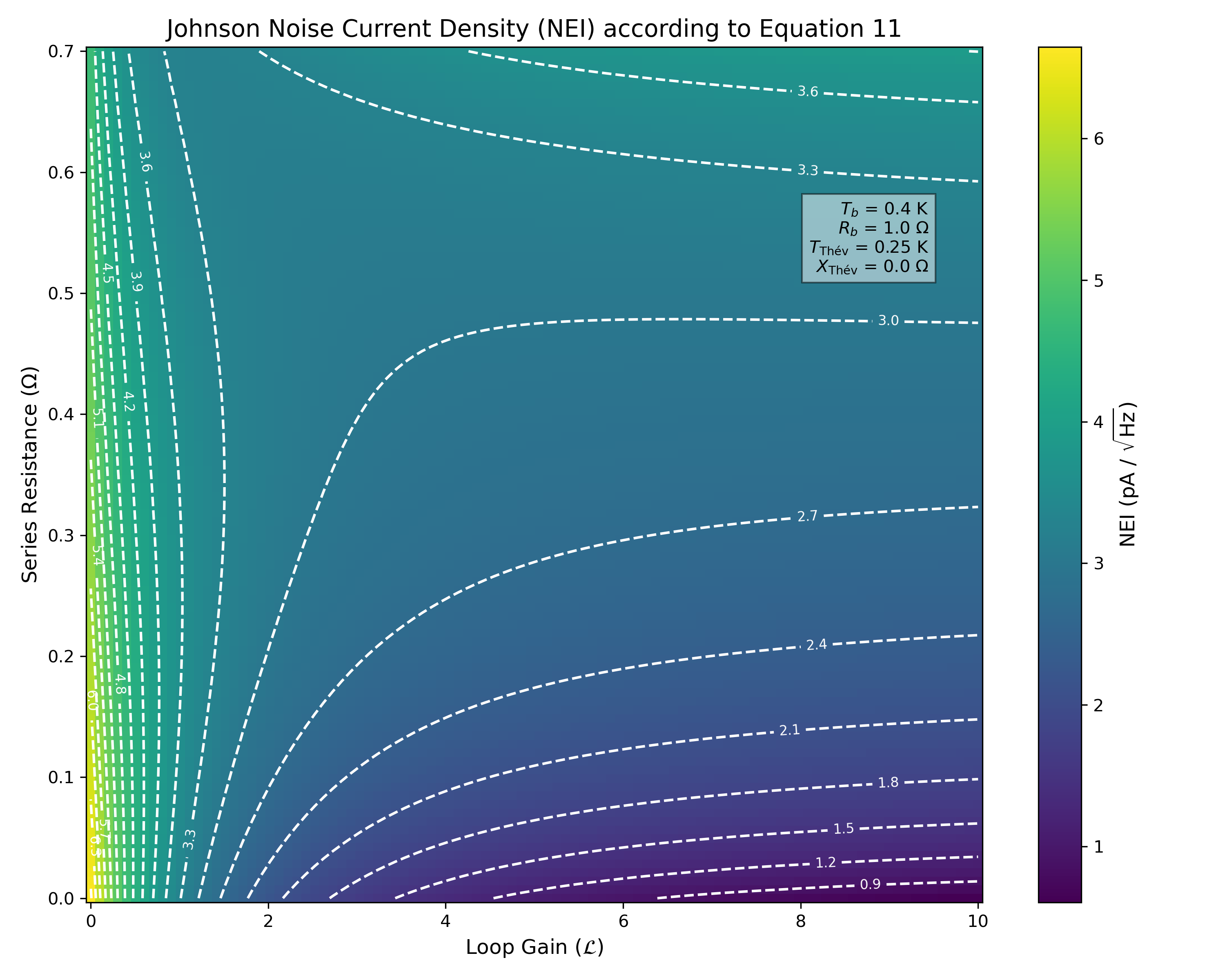}
    \caption{Johnson noise suppression as a function of electrothermal feedback loop gain $\mathcal{L}$ for representative system parameters. The general trend is for total Johnson noise to decrease with increasing loop gain and decreasing series resistance. However, at very high values of the parasitic resistance, the Johnson noise from the parasitic resistance becomes significant and the minimum NEI shifts toward $\mathcal{L}\sim1$.}
    \label{fig:jnoise}
\end{figure}

Figure~\ref{fig:jnoise} illustrates the loop-gain dependence of both noise contributions for representative parameters. Optimal Johnson noise performance is achieved with strong voltage bias (minimal parasitic impedance) and high electrothermal feedback loop gain. The analytical framework presented here has been implemented in the DfMux readout analysis software \citep{russell_tijmendfmux_calc_2025}. It enables accurate characterization of readout noise in deployed TES bolometer systems—such as SPT-3G \citep{sobrin_design_2022, montgomery_performance_2022}, Simons Array \citep{barron_integrated_2021, farias_-site_2022}—as well as design predictions for future experiments such \emph{LiteBIRD} \citep{ghigna_litebird_2024}.

\section{Conclusions}

We have presented a theoretical calculation of Johnson noise in AC-biased TES bolometers using a Thévenin-equivalent circuit representation. Our analysis demonstrates a fundamental asymmetry between Johnson noise sources located on and off the thermal island: while electrothermal feedback effectively suppresses bolometer Johnson noise by a factor of $(1+\mathcal{L})$, it provides no suppression of parasitic Johnson noise at high loop gain.

This result highlights the need for keeping the resistive parasitics low in TES readout systems. The parasitic resistance establishes a noise floor that cannot be overcome by increasing the loop gain alone. Sources of parasitic resistance such as the use of non-superconducting cables, ESR in multiplexing inductors or capacitors, and magnetic coupling to normal metals should be minimized. 

The analytical expressions derived here provide a complete description of Johnson noise in practical TES bolometer configurations with arbitrary parasitic impedances, extending previous idealized treatments. These results enable accurate noise modeling and optimization of next-generation TES arrays for cosmological and astrophysical observations.

%\section*{Acknowledgments}

\bibliography{DfMux}

%apsrev4-2.bst 2019-01-14 (MD) hand-edited version of apsrev4-1.bst
%Control: key (0)
%Control: author (8) initials jnrlst
%Control: editor formatted (1) identically to author
%Control: production of article title (0) allowed
%Control: page (0) single
%Control: year (1) truncated
%Control: production of eprint (0) enabled
\begin{thebibliography}{14}%
\makeatletter
\providecommand \@ifxundefined [1]{%
 \@ifx{#1\undefined}
}%
\providecommand \@ifnum [1]{%
 \ifnum #1\expandafter \@firstoftwo
 \else \expandafter \@secondoftwo
 \fi
}%
\providecommand \@ifx [1]{%
 \ifx #1\expandafter \@firstoftwo
 \else \expandafter \@secondoftwo
 \fi
}%
\providecommand \natexlab [1]{#1}%
\providecommand \enquote  [1]{``#1''}%
\providecommand \bibnamefont  [1]{#1}%
\providecommand \bibfnamefont [1]{#1}%
\providecommand \citenamefont [1]{#1}%
\providecommand \href@noop [0]{\@secondoftwo}%
\providecommand \href [0]{\begingroup \@sanitize@url \@href}%
\providecommand \@href[1]{\@@startlink{#1}\@@href}%
\providecommand \@@href[1]{\endgroup#1\@@endlink}%
\providecommand \@sanitize@url [0]{\catcode `\\12\catcode `\$12\catcode `\&12\catcode `\#12\catcode `\^12\catcode `\_12\catcode `\%12\relax}%
\providecommand \@@startlink[1]{}%
\providecommand \@@endlink[0]{}%
\providecommand \url  [0]{\begingroup\@sanitize@url \@url }%
\providecommand \@url [1]{\endgroup\@href {#1}{\urlprefix }}%
\providecommand \urlprefix  [0]{URL }%
\providecommand \Eprint [0]{\href }%
\providecommand \doibase [0]{https://doi.org/}%
\providecommand \selectlanguage [0]{\@gobble}%
\providecommand \bibinfo  [0]{\@secondoftwo}%
\providecommand \bibfield  [0]{\@secondoftwo}%
\providecommand \translation [1]{[#1]}%
\providecommand \BibitemOpen [0]{}%
\providecommand \bibitemStop [0]{}%
\providecommand \bibitemNoStop [0]{.\EOS\space}%
\providecommand \EOS [0]{\spacefactor3000\relax}%
\providecommand \BibitemShut  [1]{\csname bibitem#1\endcsname}%
\let\auto@bib@innerbib\@empty
%</preamble>
\bibitem [{\citenamefont {Mather}(1982)}]{mather_bolometer_1982}%
  \BibitemOpen
  \bibfield  {author} {\bibinfo {author} {\bibfnamefont {J.~C.}\ \bibnamefont {Mather}},\ }\bibfield  {title} {\bibinfo {title} {Bolometer noise: nonequilibrium theory},\ }\href {https://doi.org/10.1364/AO.21.001125} {\bibfield  {journal} {\bibinfo  {journal} {Applied Optics}\ }\textbf {\bibinfo {volume} {21}},\ \bibinfo {pages} {1125} (\bibinfo {year} {1982})},\ \bibinfo {note} {publisher: Optica Publishing Group}\BibitemShut {NoStop}%
\bibitem [{\citenamefont {Lee}\ \emph {et~al.}(1998)\citenamefont {Lee}, \citenamefont {Gildemeister}, \citenamefont {Holmes}, \citenamefont {Lee},\ and\ \citenamefont {Richards}}]{lee_voltage-biased_1998}%
  \BibitemOpen
  \bibfield  {author} {\bibinfo {author} {\bibfnamefont {S.~F.}\ \bibnamefont {Lee}}, \bibinfo {author} {\bibfnamefont {J.~M.}\ \bibnamefont {Gildemeister}}, \bibinfo {author} {\bibfnamefont {W.}~\bibnamefont {Holmes}}, \bibinfo {author} {\bibfnamefont {A.~T.}\ \bibnamefont {Lee}},\ and\ \bibinfo {author} {\bibfnamefont {P.~L.}\ \bibnamefont {Richards}},\ }\bibfield  {title} {\bibinfo {title} {Voltage-{Biased} {Superconducting} {Transition}-{Edge} {Bolometer} with {Strong} {Electrothermal} {Feedback} {Operated} at 370 {mK}},\ }\href {https://doi.org/10.1364/ao.37.003391} {\bibfield  {journal} {\bibinfo  {journal} {Applied Optics}\ }\textbf {\bibinfo {volume} {37}},\ \bibinfo {pages} {3391} (\bibinfo {year} {1998})}\BibitemShut {NoStop}%
\bibitem [{\citenamefont {Irwin}(1995)}]{irwin_application_1995}%
  \BibitemOpen
  \bibfield  {author} {\bibinfo {author} {\bibfnamefont {K.~D.}\ \bibnamefont {Irwin}},\ }\bibfield  {title} {\bibinfo {title} {An application of electrothermal feedback for high resolution cryogenic particle detection},\ }\href {https://doi.org/10.1063/1.113674} {\bibfield  {journal} {\bibinfo  {journal} {Applied Physics Letters}\ }\textbf {\bibinfo {volume} {66}},\ \bibinfo {pages} {1998} (\bibinfo {year} {1995})}\BibitemShut {NoStop}%
\bibitem [{\citenamefont {de~Wit}\ \emph {et~al.}(2021)\citenamefont {de~Wit}, \citenamefont {Gottardi}, \citenamefont {Taralli}, \citenamefont {Nagayoshi}, \citenamefont {Ridder}, \citenamefont {Akamatsu}, \citenamefont {Bruijn}, \citenamefont {Hoogeveen}, \citenamefont {van~der Kuur}, \citenamefont {Ravensberg}, \citenamefont {Vaccaro}, \citenamefont {Gao},\ and\ \citenamefont {den Herder}}]{de_wit_impact_2021}%
  \BibitemOpen
  \bibfield  {author} {\bibinfo {author} {\bibfnamefont {M.}~\bibnamefont {de~Wit}}, \bibinfo {author} {\bibfnamefont {L.}~\bibnamefont {Gottardi}}, \bibinfo {author} {\bibfnamefont {E.}~\bibnamefont {Taralli}}, \bibinfo {author} {\bibfnamefont {K.}~\bibnamefont {Nagayoshi}}, \bibinfo {author} {\bibfnamefont {M.}~\bibnamefont {Ridder}}, \bibinfo {author} {\bibfnamefont {H.}~\bibnamefont {Akamatsu}}, \bibinfo {author} {\bibfnamefont {M.}~\bibnamefont {Bruijn}}, \bibinfo {author} {\bibfnamefont {R.}~\bibnamefont {Hoogeveen}}, \bibinfo {author} {\bibfnamefont {J.}~\bibnamefont {van~der Kuur}}, \bibinfo {author} {\bibfnamefont {K.}~\bibnamefont {Ravensberg}}, \bibinfo {author} {\bibfnamefont {D.}~\bibnamefont {Vaccaro}}, \bibinfo {author} {\bibfnamefont {J.-R.}\ \bibnamefont {Gao}},\ and\ \bibinfo {author} {\bibfnamefont {J.-W.}\ \bibnamefont {den Herder}},\ }\bibfield  {title} {\bibinfo {title} {Impact of the {Absorber}-{Coupling} {Design} for {Transition}-{Edge}-{Sensor} {X}-{Ray} {Calorimeters}},\ }\href
  {https://doi.org/10.1103/PhysRevApplied.16.044059} {\bibfield  {journal} {\bibinfo  {journal} {Physical Review Applied}\ }\textbf {\bibinfo {volume} {16}},\ \bibinfo {pages} {044059} (\bibinfo {year} {2021})},\ \bibinfo {note} {publisher: American Physical Society}\BibitemShut {NoStop}%
\bibitem [{\citenamefont {Irwin}\ and\ \citenamefont {Hilton}(2005)}]{irwin_transition-edge_2005}%
  \BibitemOpen
  \bibfield  {author} {\bibinfo {author} {\bibfnamefont {K.}~\bibnamefont {Irwin}}\ and\ \bibinfo {author} {\bibfnamefont {G.}~\bibnamefont {Hilton}},\ }\bibfield  {title} {\bibinfo {title} {Transition-{Edge} {Sensors}},\ }in\ \href {https://doi.org/10.1007/10933596_3} {\emph {\bibinfo {booktitle} {Cryogenic {Particle} {Detection}}}},\ \bibinfo {editor} {edited by\ \bibinfo {editor} {\bibfnamefont {C.}~\bibnamefont {Enss}}}\ (\bibinfo  {publisher} {Springer},\ \bibinfo {address} {Berlin, Heidelberg},\ \bibinfo {year} {2005})\ pp.\ \bibinfo {pages} {63--150}\BibitemShut {NoStop}%
\bibitem [{\citenamefont {Dobbs}\ \emph {et~al.}(2012)\citenamefont {Dobbs}, \citenamefont {Lueker}, \citenamefont {Aird}, \citenamefont {Bender}, \citenamefont {Benson}, \citenamefont {Bleem}, \citenamefont {Carlstrom}, \citenamefont {Chang}, \citenamefont {Cho}, \citenamefont {Clarke}, \citenamefont {Crawford}, \citenamefont {Crites}, \citenamefont {Flanigan}, \citenamefont {de~Haan}, \citenamefont {George}, \citenamefont {Halverson}, \citenamefont {Holzapfel}, \citenamefont {Hrubes}, \citenamefont {Johnson}, \citenamefont {Joseph}, \citenamefont {Keisler}, \citenamefont {Kennedy}, \citenamefont {Kermish}, \citenamefont {Lanting}, \citenamefont {Lee}, \citenamefont {Leitch}, \citenamefont {Luong-Van}, \citenamefont {McMahon}, \citenamefont {Mehl}, \citenamefont {Meyer}, \citenamefont {Montroy}, \citenamefont {Padin}, \citenamefont {Plagge}, \citenamefont {Pryke}, \citenamefont {Richards}, \citenamefont {Ruhl}, \citenamefont {Schaffer}, \citenamefont {Schwan}, \citenamefont {Shirokoff}, \citenamefont
  {Spieler}, \citenamefont {Staniszewski}, \citenamefont {Stark}, \citenamefont {Vanderlinde}, \citenamefont {Vieira}, \citenamefont {Vu}, \citenamefont {Westbrook},\ and\ \citenamefont {Williamson}}]{dobbs_frequency_2012}%
  \BibitemOpen
  \bibfield  {author} {\bibinfo {author} {\bibfnamefont {M.~A.}\ \bibnamefont {Dobbs}}, \bibinfo {author} {\bibfnamefont {M.}~\bibnamefont {Lueker}}, \bibinfo {author} {\bibfnamefont {K.~A.}\ \bibnamefont {Aird}}, \bibinfo {author} {\bibfnamefont {A.~N.}\ \bibnamefont {Bender}}, \bibinfo {author} {\bibfnamefont {B.~A.}\ \bibnamefont {Benson}}, \bibinfo {author} {\bibfnamefont {L.~E.}\ \bibnamefont {Bleem}}, \bibinfo {author} {\bibfnamefont {J.~E.}\ \bibnamefont {Carlstrom}}, \bibinfo {author} {\bibfnamefont {C.~L.}\ \bibnamefont {Chang}}, \bibinfo {author} {\bibfnamefont {H.-M.}\ \bibnamefont {Cho}}, \bibinfo {author} {\bibfnamefont {J.}~\bibnamefont {Clarke}}, \bibinfo {author} {\bibfnamefont {T.~M.}\ \bibnamefont {Crawford}}, \bibinfo {author} {\bibfnamefont {A.~T.}\ \bibnamefont {Crites}}, \bibinfo {author} {\bibfnamefont {D.~I.}\ \bibnamefont {Flanigan}}, \bibinfo {author} {\bibfnamefont {T.}~\bibnamefont {de~Haan}}, \bibinfo {author} {\bibfnamefont {E.~M.}\ \bibnamefont {George}}, \bibinfo {author}
  {\bibfnamefont {N.~W.}\ \bibnamefont {Halverson}}, \bibinfo {author} {\bibfnamefont {W.~L.}\ \bibnamefont {Holzapfel}}, \bibinfo {author} {\bibfnamefont {J.~D.}\ \bibnamefont {Hrubes}}, \bibinfo {author} {\bibfnamefont {B.~R.}\ \bibnamefont {Johnson}}, \bibinfo {author} {\bibfnamefont {J.}~\bibnamefont {Joseph}}, \bibinfo {author} {\bibfnamefont {R.}~\bibnamefont {Keisler}}, \bibinfo {author} {\bibfnamefont {J.}~\bibnamefont {Kennedy}}, \bibinfo {author} {\bibfnamefont {Z.}~\bibnamefont {Kermish}}, \bibinfo {author} {\bibfnamefont {T.~M.}\ \bibnamefont {Lanting}}, \bibinfo {author} {\bibfnamefont {A.~T.}\ \bibnamefont {Lee}}, \bibinfo {author} {\bibfnamefont {E.~M.}\ \bibnamefont {Leitch}}, \bibinfo {author} {\bibfnamefont {D.}~\bibnamefont {Luong-Van}}, \bibinfo {author} {\bibfnamefont {J.~J.}\ \bibnamefont {McMahon}}, \bibinfo {author} {\bibfnamefont {J.}~\bibnamefont {Mehl}}, \bibinfo {author} {\bibfnamefont {S.~S.}\ \bibnamefont {Meyer}}, \bibinfo {author} {\bibfnamefont {T.~E.}\ \bibnamefont
  {Montroy}}, \bibinfo {author} {\bibfnamefont {S.}~\bibnamefont {Padin}}, \bibinfo {author} {\bibfnamefont {T.}~\bibnamefont {Plagge}}, \bibinfo {author} {\bibfnamefont {C.}~\bibnamefont {Pryke}}, \bibinfo {author} {\bibfnamefont {P.~L.}\ \bibnamefont {Richards}}, \bibinfo {author} {\bibfnamefont {J.~E.}\ \bibnamefont {Ruhl}}, \bibinfo {author} {\bibfnamefont {K.~K.}\ \bibnamefont {Schaffer}}, \bibinfo {author} {\bibfnamefont {D.}~\bibnamefont {Schwan}}, \bibinfo {author} {\bibfnamefont {E.}~\bibnamefont {Shirokoff}}, \bibinfo {author} {\bibfnamefont {H.~G.}\ \bibnamefont {Spieler}}, \bibinfo {author} {\bibfnamefont {Z.}~\bibnamefont {Staniszewski}}, \bibinfo {author} {\bibfnamefont {A.~A.}\ \bibnamefont {Stark}}, \bibinfo {author} {\bibfnamefont {K.}~\bibnamefont {Vanderlinde}}, \bibinfo {author} {\bibfnamefont {J.~D.}\ \bibnamefont {Vieira}}, \bibinfo {author} {\bibfnamefont {C.}~\bibnamefont {Vu}}, \bibinfo {author} {\bibfnamefont {B.}~\bibnamefont {Westbrook}},\ and\ \bibinfo {author} {\bibfnamefont
  {R.}~\bibnamefont {Williamson}},\ }\bibfield  {title} {\bibinfo {title} {Frequency multiplexed superconducting quantum interference device readout of large bolometer arrays for cosmic microwave background measurements},\ }\href {https://doi.org/10.1063/1.4737629} {\bibfield  {journal} {\bibinfo  {journal} {Review of Scientific Instruments}\ }\textbf {\bibinfo {volume} {83}},\ \bibinfo {pages} {073113} (\bibinfo {year} {2012})}\BibitemShut {NoStop}%
\bibitem [{\citenamefont {de~Haan}(2024)}]{de_haan_mntes_2024}%
  \BibitemOpen
  \bibfield  {author} {\bibinfo {author} {\bibfnamefont {T.}~\bibnamefont {de~Haan}},\ }\bibfield  {title} {\bibinfo {title} {{MNTES}: modeling nonlinearity of {TES} detectors for enhanced cosmic microwave background measurements with {LiteBIRD}},\ }in\ \href {https://doi.org/10.1117/12.3018503} {\emph {\bibinfo {booktitle} {Millimeter, {Submillimeter}, and {Far}-{Infrared} {Detectors} and {Instrumentation} for {Astronomy} {XII}}}},\ Vol.\ \bibinfo {volume} {13102},\ \bibinfo {editor} {edited by\ \bibinfo {editor} {\bibfnamefont {J.}~\bibnamefont {Zmuidzinas}}\ and\ \bibinfo {editor} {\bibfnamefont {J.-R.}\ \bibnamefont {Gao}}}\ (\bibinfo  {publisher} {SPIE},\ \bibinfo {year} {2024})\ p.\ \bibinfo {pages} {1310208},\ \bibinfo {note} {backup Publisher: International Society for Optics and Photonics}\BibitemShut {NoStop}%
\bibitem [{\citenamefont {Zhou}\ \emph {et~al.}(2024)\citenamefont {Zhou}, \citenamefont {de~Haan}, \citenamefont {Akamatsu}, \citenamefont {Kaneko}, \citenamefont {Hazumi}, \citenamefont {Hasegawa}, \citenamefont {Suzuki},\ and\ \citenamefont {Lee}}]{zhou_method_2024}%
  \BibitemOpen
  \bibfield  {author} {\bibinfo {author} {\bibfnamefont {Y.}~\bibnamefont {Zhou}}, \bibinfo {author} {\bibfnamefont {T.}~\bibnamefont {de~Haan}}, \bibinfo {author} {\bibfnamefont {H.}~\bibnamefont {Akamatsu}}, \bibinfo {author} {\bibfnamefont {D.}~\bibnamefont {Kaneko}}, \bibinfo {author} {\bibfnamefont {M.}~\bibnamefont {Hazumi}}, \bibinfo {author} {\bibfnamefont {M.}~\bibnamefont {Hasegawa}}, \bibinfo {author} {\bibfnamefont {A.}~\bibnamefont {Suzuki}},\ and\ \bibinfo {author} {\bibfnamefont {A.~T.}\ \bibnamefont {Lee}},\ }\bibfield  {title} {\bibinfo {title} {A {Method} of {Measuring} {TES} {Complex} {ETF} {Response} in {Frequency}-{Domain} {Multiplexed} {Readout} by {Single} {Sideband} {Power} {Modulation}},\ }\bibfield  {journal} {\bibinfo  {journal} {Journal of Low Temperature Physics}\ }\href {https://doi.org/10.1007/s10909-024-03107-z} {10.1007/s10909-024-03107-z} (\bibinfo {year} {2024})\BibitemShut {NoStop}%
\bibitem [{\citenamefont {Russell}\ \emph {et~al.}(2025)\citenamefont {Russell}, \citenamefont {de~Haan},\ and\ \citenamefont {Farias}}]{russell_tijmendfmux_calc_2025}%
  \BibitemOpen
  \bibfield  {author} {\bibinfo {author} {\bibfnamefont {M.}~\bibnamefont {Russell}}, \bibinfo {author} {\bibfnamefont {T.}~\bibnamefont {de~Haan}},\ and\ \bibinfo {author} {\bibfnamefont {N.}~\bibnamefont {Farias}},\ }\href {https://github.com/tijmen/dfmux_calc} {\bibinfo {title} {tijmen/dfmux\_calc}} (\bibinfo {year} {2025}),\ \bibinfo {note} {original-date: 2024-03-05T02:42:11Z}\BibitemShut {NoStop}%
\bibitem [{\citenamefont {Sobrin}\ \emph {et~al.}(2022)\citenamefont {Sobrin}, \citenamefont {Anderson}, \citenamefont {Bender}, \citenamefont {Benson}, \citenamefont {Dutcher}, \citenamefont {Foster}, \citenamefont {Goeckner-Wald}, \citenamefont {Montgomery}, \citenamefont {Nadolski}, \citenamefont {Rahlin}, \citenamefont {Ade}, \citenamefont {Ahmed}, \citenamefont {Anderes}, \citenamefont {Archipley}, \citenamefont {Austermann}, \citenamefont {Avva}, \citenamefont {Aylor}, \citenamefont {Balkenhol}, \citenamefont {Barry}, \citenamefont {Thakur}, \citenamefont {Benabed}, \citenamefont {Bianchini}, \citenamefont {Bleem}, \citenamefont {Bouchet}, \citenamefont {Bryant}, \citenamefont {Byrum}, \citenamefont {Carlstrom}, \citenamefont {Carter}, \citenamefont {Cecil}, \citenamefont {Chang}, \citenamefont {Chaubal}, \citenamefont {Chen}, \citenamefont {Cho}, \citenamefont {Chou}, \citenamefont {Cliche}, \citenamefont {Crawford}, \citenamefont {Cukierman}, \citenamefont {Daley}, \citenamefont {de~Haan},
  \citenamefont {Denison}, \citenamefont {Dibert}, \citenamefont {Ding}, \citenamefont {Dobbs}, \citenamefont {Everett}, \citenamefont {Feng}, \citenamefont {Ferguson}, \citenamefont {Fu}, \citenamefont {Galli}, \citenamefont {Gambrel}, \citenamefont {Gardner}, \citenamefont {Gualtieri}, \citenamefont {Guns}, \citenamefont {Gupta}, \citenamefont {Guyser}, \citenamefont {Halverson}, \citenamefont {Harke-Hosemann}, \citenamefont {Harrington}, \citenamefont {Henning}, \citenamefont {Hilton}, \citenamefont {Hivon}, \citenamefont {Holder}, \citenamefont {Holzapfel}, \citenamefont {Hood}, \citenamefont {Howe}, \citenamefont {Huang}, \citenamefont {Irwin}, \citenamefont {Jeong}, \citenamefont {Jonas}, \citenamefont {Jones}, \citenamefont {Khaire}, \citenamefont {Knox}, \citenamefont {Kofman}, \citenamefont {Korman}, \citenamefont {Kubik}, \citenamefont {Kuhlmann}, \citenamefont {Kuo}, \citenamefont {Lee}, \citenamefont {Leitch}, \citenamefont {Lowitz}, \citenamefont {Lu}, \citenamefont {Meyer}, \citenamefont
  {Michalik}, \citenamefont {Millea}, \citenamefont {Natoli}, \citenamefont {Nguyen}, \citenamefont {Noble}, \citenamefont {Novosad}, \citenamefont {Omori}, \citenamefont {Padin}, \citenamefont {Pan}, \citenamefont {Paschos}, \citenamefont {Pearson}, \citenamefont {Posada}, \citenamefont {Prabhu}, \citenamefont {Quan}, \citenamefont {Reichardt}, \citenamefont {Riebel}, \citenamefont {Riedel}, \citenamefont {Rouble}, \citenamefont {Ruhl}, \citenamefont {Saliwanchik}, \citenamefont {Sayre}, \citenamefont {Schiappucci}, \citenamefont {Shirokoff}, \citenamefont {Smecher}, \citenamefont {Stark}, \citenamefont {Stephen}, \citenamefont {Story}, \citenamefont {Suzuki}, \citenamefont {Tandoi}, \citenamefont {Thompson}, \citenamefont {Thorne}, \citenamefont {Tucker}, \citenamefont {Umilta}, \citenamefont {Vale}, \citenamefont {Vanderlinde}, \citenamefont {Vieira}, \citenamefont {Wang}, \citenamefont {Whitehorn}, \citenamefont {Wu}, \citenamefont {Yefremenko}, \citenamefont {Yoon},\ and\ \citenamefont
  {Young}}]{sobrin_design_2022}%
  \BibitemOpen
  \bibfield  {author} {\bibinfo {author} {\bibfnamefont {J.~A.}\ \bibnamefont {Sobrin}}, \bibinfo {author} {\bibfnamefont {A.~J.}\ \bibnamefont {Anderson}}, \bibinfo {author} {\bibfnamefont {A.~N.}\ \bibnamefont {Bender}}, \bibinfo {author} {\bibfnamefont {B.~A.}\ \bibnamefont {Benson}}, \bibinfo {author} {\bibfnamefont {D.}~\bibnamefont {Dutcher}}, \bibinfo {author} {\bibfnamefont {A.}~\bibnamefont {Foster}}, \bibinfo {author} {\bibfnamefont {N.}~\bibnamefont {Goeckner-Wald}}, \bibinfo {author} {\bibfnamefont {J.}~\bibnamefont {Montgomery}}, \bibinfo {author} {\bibfnamefont {A.}~\bibnamefont {Nadolski}}, \bibinfo {author} {\bibfnamefont {A.}~\bibnamefont {Rahlin}}, \bibinfo {author} {\bibfnamefont {P.~A.~R.}\ \bibnamefont {Ade}}, \bibinfo {author} {\bibfnamefont {Z.}~\bibnamefont {Ahmed}}, \bibinfo {author} {\bibfnamefont {E.}~\bibnamefont {Anderes}}, \bibinfo {author} {\bibfnamefont {M.}~\bibnamefont {Archipley}}, \bibinfo {author} {\bibfnamefont {J.~E.}\ \bibnamefont {Austermann}}, \bibinfo {author}
  {\bibfnamefont {J.~S.}\ \bibnamefont {Avva}}, \bibinfo {author} {\bibfnamefont {K.}~\bibnamefont {Aylor}}, \bibinfo {author} {\bibfnamefont {L.}~\bibnamefont {Balkenhol}}, \bibinfo {author} {\bibfnamefont {P.~S.}\ \bibnamefont {Barry}}, \bibinfo {author} {\bibfnamefont {R.~B.}\ \bibnamefont {Thakur}}, \bibinfo {author} {\bibfnamefont {K.}~\bibnamefont {Benabed}}, \bibinfo {author} {\bibfnamefont {F.}~\bibnamefont {Bianchini}}, \bibinfo {author} {\bibfnamefont {L.~E.}\ \bibnamefont {Bleem}}, \bibinfo {author} {\bibfnamefont {F.~R.}\ \bibnamefont {Bouchet}}, \bibinfo {author} {\bibfnamefont {L.}~\bibnamefont {Bryant}}, \bibinfo {author} {\bibfnamefont {K.}~\bibnamefont {Byrum}}, \bibinfo {author} {\bibfnamefont {J.~E.}\ \bibnamefont {Carlstrom}}, \bibinfo {author} {\bibfnamefont {F.~W.}\ \bibnamefont {Carter}}, \bibinfo {author} {\bibfnamefont {T.~W.}\ \bibnamefont {Cecil}}, \bibinfo {author} {\bibfnamefont {C.~L.}\ \bibnamefont {Chang}}, \bibinfo {author} {\bibfnamefont {P.}~\bibnamefont {Chaubal}}, \bibinfo
  {author} {\bibfnamefont {G.}~\bibnamefont {Chen}}, \bibinfo {author} {\bibfnamefont {H.-M.}\ \bibnamefont {Cho}}, \bibinfo {author} {\bibfnamefont {T.-L.}\ \bibnamefont {Chou}}, \bibinfo {author} {\bibfnamefont {J.-F.}\ \bibnamefont {Cliche}}, \bibinfo {author} {\bibfnamefont {T.~M.}\ \bibnamefont {Crawford}}, \bibinfo {author} {\bibfnamefont {A.}~\bibnamefont {Cukierman}}, \bibinfo {author} {\bibfnamefont {C.}~\bibnamefont {Daley}}, \bibinfo {author} {\bibfnamefont {T.}~\bibnamefont {de~Haan}}, \bibinfo {author} {\bibfnamefont {E.~V.}\ \bibnamefont {Denison}}, \bibinfo {author} {\bibfnamefont {K.}~\bibnamefont {Dibert}}, \bibinfo {author} {\bibfnamefont {J.}~\bibnamefont {Ding}}, \bibinfo {author} {\bibfnamefont {M.~A.}\ \bibnamefont {Dobbs}}, \bibinfo {author} {\bibfnamefont {W.}~\bibnamefont {Everett}}, \bibinfo {author} {\bibfnamefont {C.}~\bibnamefont {Feng}}, \bibinfo {author} {\bibfnamefont {K.~R.}\ \bibnamefont {Ferguson}}, \bibinfo {author} {\bibfnamefont {J.}~\bibnamefont {Fu}}, \bibinfo {author}
  {\bibfnamefont {S.}~\bibnamefont {Galli}}, \bibinfo {author} {\bibfnamefont {A.~E.}\ \bibnamefont {Gambrel}}, \bibinfo {author} {\bibfnamefont {R.~W.}\ \bibnamefont {Gardner}}, \bibinfo {author} {\bibfnamefont {R.}~\bibnamefont {Gualtieri}}, \bibinfo {author} {\bibfnamefont {S.}~\bibnamefont {Guns}}, \bibinfo {author} {\bibfnamefont {N.}~\bibnamefont {Gupta}}, \bibinfo {author} {\bibfnamefont {R.}~\bibnamefont {Guyser}}, \bibinfo {author} {\bibfnamefont {N.~W.}\ \bibnamefont {Halverson}}, \bibinfo {author} {\bibfnamefont {A.~H.}\ \bibnamefont {Harke-Hosemann}}, \bibinfo {author} {\bibfnamefont {N.~L.}\ \bibnamefont {Harrington}}, \bibinfo {author} {\bibfnamefont {J.~W.}\ \bibnamefont {Henning}}, \bibinfo {author} {\bibfnamefont {G.~C.}\ \bibnamefont {Hilton}}, \bibinfo {author} {\bibfnamefont {E.}~\bibnamefont {Hivon}}, \bibinfo {author} {\bibfnamefont {G.~P.}\ \bibnamefont {Holder}}, \bibinfo {author} {\bibfnamefont {W.~L.}\ \bibnamefont {Holzapfel}}, \bibinfo {author} {\bibfnamefont {J.~C.}\ \bibnamefont
  {Hood}}, \bibinfo {author} {\bibfnamefont {D.}~\bibnamefont {Howe}}, \bibinfo {author} {\bibfnamefont {N.}~\bibnamefont {Huang}}, \bibinfo {author} {\bibfnamefont {K.~D.}\ \bibnamefont {Irwin}}, \bibinfo {author} {\bibfnamefont {O.~B.}\ \bibnamefont {Jeong}}, \bibinfo {author} {\bibfnamefont {M.}~\bibnamefont {Jonas}}, \bibinfo {author} {\bibfnamefont {A.}~\bibnamefont {Jones}}, \bibinfo {author} {\bibfnamefont {T.~S.}\ \bibnamefont {Khaire}}, \bibinfo {author} {\bibfnamefont {L.}~\bibnamefont {Knox}}, \bibinfo {author} {\bibfnamefont {A.~M.}\ \bibnamefont {Kofman}}, \bibinfo {author} {\bibfnamefont {M.}~\bibnamefont {Korman}}, \bibinfo {author} {\bibfnamefont {D.~L.}\ \bibnamefont {Kubik}}, \bibinfo {author} {\bibfnamefont {S.}~\bibnamefont {Kuhlmann}}, \bibinfo {author} {\bibfnamefont {C.-L.}\ \bibnamefont {Kuo}}, \bibinfo {author} {\bibfnamefont {A.~T.}\ \bibnamefont {Lee}}, \bibinfo {author} {\bibfnamefont {E.~M.}\ \bibnamefont {Leitch}}, \bibinfo {author} {\bibfnamefont {A.~E.}\ \bibnamefont {Lowitz}},
  \bibinfo {author} {\bibfnamefont {C.}~\bibnamefont {Lu}}, \bibinfo {author} {\bibfnamefont {S.~S.}\ \bibnamefont {Meyer}}, \bibinfo {author} {\bibfnamefont {D.}~\bibnamefont {Michalik}}, \bibinfo {author} {\bibfnamefont {M.}~\bibnamefont {Millea}}, \bibinfo {author} {\bibfnamefont {T.}~\bibnamefont {Natoli}}, \bibinfo {author} {\bibfnamefont {H.}~\bibnamefont {Nguyen}}, \bibinfo {author} {\bibfnamefont {G.~I.}\ \bibnamefont {Noble}}, \bibinfo {author} {\bibfnamefont {V.}~\bibnamefont {Novosad}}, \bibinfo {author} {\bibfnamefont {Y.}~\bibnamefont {Omori}}, \bibinfo {author} {\bibfnamefont {S.}~\bibnamefont {Padin}}, \bibinfo {author} {\bibfnamefont {Z.}~\bibnamefont {Pan}}, \bibinfo {author} {\bibfnamefont {P.}~\bibnamefont {Paschos}}, \bibinfo {author} {\bibfnamefont {J.}~\bibnamefont {Pearson}}, \bibinfo {author} {\bibfnamefont {C.~M.}\ \bibnamefont {Posada}}, \bibinfo {author} {\bibfnamefont {K.}~\bibnamefont {Prabhu}}, \bibinfo {author} {\bibfnamefont {W.}~\bibnamefont {Quan}}, \bibinfo {author}
  {\bibfnamefont {C.~L.}\ \bibnamefont {Reichardt}}, \bibinfo {author} {\bibfnamefont {D.}~\bibnamefont {Riebel}}, \bibinfo {author} {\bibfnamefont {B.}~\bibnamefont {Riedel}}, \bibinfo {author} {\bibfnamefont {M.}~\bibnamefont {Rouble}}, \bibinfo {author} {\bibfnamefont {J.~E.}\ \bibnamefont {Ruhl}}, \bibinfo {author} {\bibfnamefont {B.}~\bibnamefont {Saliwanchik}}, \bibinfo {author} {\bibfnamefont {J.~T.}\ \bibnamefont {Sayre}}, \bibinfo {author} {\bibfnamefont {E.}~\bibnamefont {Schiappucci}}, \bibinfo {author} {\bibfnamefont {E.}~\bibnamefont {Shirokoff}}, \bibinfo {author} {\bibfnamefont {G.}~\bibnamefont {Smecher}}, \bibinfo {author} {\bibfnamefont {A.~A.}\ \bibnamefont {Stark}}, \bibinfo {author} {\bibfnamefont {J.}~\bibnamefont {Stephen}}, \bibinfo {author} {\bibfnamefont {K.~T.}\ \bibnamefont {Story}}, \bibinfo {author} {\bibfnamefont {A.}~\bibnamefont {Suzuki}}, \bibinfo {author} {\bibfnamefont {C.}~\bibnamefont {Tandoi}}, \bibinfo {author} {\bibfnamefont {K.~L.}\ \bibnamefont {Thompson}}, \bibinfo
  {author} {\bibfnamefont {B.}~\bibnamefont {Thorne}}, \bibinfo {author} {\bibfnamefont {C.}~\bibnamefont {Tucker}}, \bibinfo {author} {\bibfnamefont {C.}~\bibnamefont {Umilta}}, \bibinfo {author} {\bibfnamefont {L.~R.}\ \bibnamefont {Vale}}, \bibinfo {author} {\bibfnamefont {K.}~\bibnamefont {Vanderlinde}}, \bibinfo {author} {\bibfnamefont {J.~D.}\ \bibnamefont {Vieira}}, \bibinfo {author} {\bibfnamefont {G.}~\bibnamefont {Wang}}, \bibinfo {author} {\bibfnamefont {N.}~\bibnamefont {Whitehorn}}, \bibinfo {author} {\bibfnamefont {W.~L.~K.}\ \bibnamefont {Wu}}, \bibinfo {author} {\bibfnamefont {V.}~\bibnamefont {Yefremenko}}, \bibinfo {author} {\bibfnamefont {K.~W.}\ \bibnamefont {Yoon}},\ and\ \bibinfo {author} {\bibfnamefont {M.~R.}\ \bibnamefont {Young}},\ }\bibfield  {title} {\bibinfo {title} {The {Design} and {Integrated} {Performance} of {SPT}-{3G}},\ }\href {https://doi.org/10.3847/1538-4365/ac374f} {\bibfield  {journal} {\bibinfo  {journal} {The Astrophysical Journal Supplement Series}\ }\textbf
  {\bibinfo {volume} {258}},\ \bibinfo {pages} {42} (\bibinfo {year} {2022})},\ \bibinfo {note} {arXiv:2106.11202 [astro-ph]}\BibitemShut {NoStop}%
\bibitem [{\citenamefont {Montgomery}\ \emph {et~al.}(2022)\citenamefont {Montgomery}, \citenamefont {Ade}, \citenamefont {Ahmed}, \citenamefont {Anderes}, \citenamefont {Anderson}, \citenamefont {Archipley}, \citenamefont {Avva}, \citenamefont {Aylor}, \citenamefont {Balkenhol}, \citenamefont {Barry}, \citenamefont {Thakur}, \citenamefont {Benabed}, \citenamefont {Bender}, \citenamefont {Benson}, \citenamefont {Bianchini}, \citenamefont {Bleem}, \citenamefont {Bouchet}, \citenamefont {Bryant}, \citenamefont {Byrum}, \citenamefont {Carlstrom}, \citenamefont {Carter}, \citenamefont {Cecil}, \citenamefont {Chang}, \citenamefont {Chaubal}, \citenamefont {Chen}, \citenamefont {Cho}, \citenamefont {Chou}, \citenamefont {Cliche}, \citenamefont {Crawford}, \citenamefont {Cukierman}, \citenamefont {Daley}, \citenamefont {de~Haan}, \citenamefont {Denison}, \citenamefont {Dibert}, \citenamefont {Ding}, \citenamefont {Dobbs}, \citenamefont {Dutcher}, \citenamefont {Elleflot}, \citenamefont {Everett}, \citenamefont
  {Feng}, \citenamefont {Ferguson}, \citenamefont {Foster}, \citenamefont {Fu}, \citenamefont {Galli}, \citenamefont {Gambrel}, \citenamefont {Gardner}, \citenamefont {Goeckner-Wald}, \citenamefont {Groh}, \citenamefont {Gualtieri}, \citenamefont {Guns}, \citenamefont {Gupta}, \citenamefont {Guyser}, \citenamefont {Halverson}, \citenamefont {Harke-Hosemann}, \citenamefont {Harrington}, \citenamefont {Henning}, \citenamefont {Hilton}, \citenamefont {Hivon}, \citenamefont {Holzapfel}, \citenamefont {Hood}, \citenamefont {Howe}, \citenamefont {Huang}, \citenamefont {Irwin}, \citenamefont {Jeong}, \citenamefont {Jonas}, \citenamefont {Jones}, \citenamefont {Khaire}, \citenamefont {Knox}, \citenamefont {Kofman}, \citenamefont {Korman}, \citenamefont {Kubik}, \citenamefont {Kuhlmann}, \citenamefont {Kuo}, \citenamefont {Lee}, \citenamefont {Leitch}, \citenamefont {Lowitz}, \citenamefont {Lu}, \citenamefont {Meyer}, \citenamefont {Michalik}, \citenamefont {Millea}, \citenamefont {Nadolski}, \citenamefont {Natoli},
  \citenamefont {Nguyen}, \citenamefont {Noble}, \citenamefont {Novosad}, \citenamefont {Omori}, \citenamefont {Padin}, \citenamefont {Pan}, \citenamefont {Paschos}, \citenamefont {Pearson}, \citenamefont {Posada}, \citenamefont {Prabhu}, \citenamefont {Quan}, \citenamefont {Rahlin}, \citenamefont {Reichardt}, \citenamefont {Riebel}, \citenamefont {Riedel}, \citenamefont {Rouble}, \citenamefont {Ruhl}, \citenamefont {Sayre}, \citenamefont {Schiappucci}, \citenamefont {Shirokoff}, \citenamefont {Smecher}, \citenamefont {Sobrin}, \citenamefont {Stark}, \citenamefont {Stephen}, \citenamefont {Story}, \citenamefont {Suzuki}, \citenamefont {Thompson}, \citenamefont {Thorne}, \citenamefont {Tucker}, \citenamefont {Umilta}, \citenamefont {Vale}, \citenamefont {Vanderlinde}, \citenamefont {Vieira}, \citenamefont {Wang}, \citenamefont {Whitehorn}, \citenamefont {Wu}, \citenamefont {Yefremenko}, \citenamefont {Yoon},\ and\ \citenamefont {Young}}]{montgomery_performance_2022}%
  \BibitemOpen
  \bibfield  {author} {\bibinfo {author} {\bibfnamefont {J.}~\bibnamefont {Montgomery}}, \bibinfo {author} {\bibfnamefont {P.~A.~R.}\ \bibnamefont {Ade}}, \bibinfo {author} {\bibfnamefont {Z.}~\bibnamefont {Ahmed}}, \bibinfo {author} {\bibfnamefont {E.}~\bibnamefont {Anderes}}, \bibinfo {author} {\bibfnamefont {A.~J.}\ \bibnamefont {Anderson}}, \bibinfo {author} {\bibfnamefont {M.}~\bibnamefont {Archipley}}, \bibinfo {author} {\bibfnamefont {J.~S.}\ \bibnamefont {Avva}}, \bibinfo {author} {\bibfnamefont {K.}~\bibnamefont {Aylor}}, \bibinfo {author} {\bibfnamefont {L.}~\bibnamefont {Balkenhol}}, \bibinfo {author} {\bibfnamefont {P.~S.}\ \bibnamefont {Barry}}, \bibinfo {author} {\bibfnamefont {R.~B.}\ \bibnamefont {Thakur}}, \bibinfo {author} {\bibfnamefont {K.}~\bibnamefont {Benabed}}, \bibinfo {author} {\bibfnamefont {A.~N.}\ \bibnamefont {Bender}}, \bibinfo {author} {\bibfnamefont {B.~A.}\ \bibnamefont {Benson}}, \bibinfo {author} {\bibfnamefont {F.}~\bibnamefont {Bianchini}}, \bibinfo {author}
  {\bibfnamefont {L.~E.}\ \bibnamefont {Bleem}}, \bibinfo {author} {\bibfnamefont {F.~R.}\ \bibnamefont {Bouchet}}, \bibinfo {author} {\bibfnamefont {L.}~\bibnamefont {Bryant}}, \bibinfo {author} {\bibfnamefont {K.}~\bibnamefont {Byrum}}, \bibinfo {author} {\bibfnamefont {J.~E.}\ \bibnamefont {Carlstrom}}, \bibinfo {author} {\bibfnamefont {F.~W.}\ \bibnamefont {Carter}}, \bibinfo {author} {\bibfnamefont {T.~W.}\ \bibnamefont {Cecil}}, \bibinfo {author} {\bibfnamefont {C.~L.}\ \bibnamefont {Chang}}, \bibinfo {author} {\bibfnamefont {P.}~\bibnamefont {Chaubal}}, \bibinfo {author} {\bibfnamefont {G.}~\bibnamefont {Chen}}, \bibinfo {author} {\bibfnamefont {H.-M.}\ \bibnamefont {Cho}}, \bibinfo {author} {\bibfnamefont {T.-L.}\ \bibnamefont {Chou}}, \bibinfo {author} {\bibfnamefont {J.-F.}\ \bibnamefont {Cliche}}, \bibinfo {author} {\bibfnamefont {T.~M.}\ \bibnamefont {Crawford}}, \bibinfo {author} {\bibfnamefont {A.}~\bibnamefont {Cukierman}}, \bibinfo {author} {\bibfnamefont {C.}~\bibnamefont {Daley}}, \bibinfo
  {author} {\bibfnamefont {T.}~\bibnamefont {de~Haan}}, \bibinfo {author} {\bibfnamefont {E.~V.}\ \bibnamefont {Denison}}, \bibinfo {author} {\bibfnamefont {K.}~\bibnamefont {Dibert}}, \bibinfo {author} {\bibfnamefont {J.}~\bibnamefont {Ding}}, \bibinfo {author} {\bibfnamefont {M.~A.}\ \bibnamefont {Dobbs}}, \bibinfo {author} {\bibfnamefont {D.}~\bibnamefont {Dutcher}}, \bibinfo {author} {\bibfnamefont {T.}~\bibnamefont {Elleflot}}, \bibinfo {author} {\bibfnamefont {W.}~\bibnamefont {Everett}}, \bibinfo {author} {\bibfnamefont {C.}~\bibnamefont {Feng}}, \bibinfo {author} {\bibfnamefont {K.~R.}\ \bibnamefont {Ferguson}}, \bibinfo {author} {\bibfnamefont {A.}~\bibnamefont {Foster}}, \bibinfo {author} {\bibfnamefont {J.}~\bibnamefont {Fu}}, \bibinfo {author} {\bibfnamefont {S.}~\bibnamefont {Galli}}, \bibinfo {author} {\bibfnamefont {A.~E.}\ \bibnamefont {Gambrel}}, \bibinfo {author} {\bibfnamefont {R.~W.}\ \bibnamefont {Gardner}}, \bibinfo {author} {\bibfnamefont {N.}~\bibnamefont {Goeckner-Wald}}, \bibinfo
  {author} {\bibfnamefont {J.~C.}\ \bibnamefont {Groh}}, \bibinfo {author} {\bibfnamefont {R.}~\bibnamefont {Gualtieri}}, \bibinfo {author} {\bibfnamefont {S.}~\bibnamefont {Guns}}, \bibinfo {author} {\bibfnamefont {N.}~\bibnamefont {Gupta}}, \bibinfo {author} {\bibfnamefont {R.}~\bibnamefont {Guyser}}, \bibinfo {author} {\bibfnamefont {N.~W.}\ \bibnamefont {Halverson}}, \bibinfo {author} {\bibfnamefont {A.~H.}\ \bibnamefont {Harke-Hosemann}}, \bibinfo {author} {\bibfnamefont {N.~L.}\ \bibnamefont {Harrington}}, \bibinfo {author} {\bibfnamefont {J.~W.}\ \bibnamefont {Henning}}, \bibinfo {author} {\bibfnamefont {G.~C.}\ \bibnamefont {Hilton}}, \bibinfo {author} {\bibfnamefont {E.}~\bibnamefont {Hivon}}, \bibinfo {author} {\bibfnamefont {W.~L.}\ \bibnamefont {Holzapfel}}, \bibinfo {author} {\bibfnamefont {J.~C.}\ \bibnamefont {Hood}}, \bibinfo {author} {\bibfnamefont {D.}~\bibnamefont {Howe}}, \bibinfo {author} {\bibfnamefont {N.}~\bibnamefont {Huang}}, \bibinfo {author} {\bibfnamefont {K.~D.}\ \bibnamefont
  {Irwin}}, \bibinfo {author} {\bibfnamefont {O.~B.}\ \bibnamefont {Jeong}}, \bibinfo {author} {\bibfnamefont {M.}~\bibnamefont {Jonas}}, \bibinfo {author} {\bibfnamefont {A.}~\bibnamefont {Jones}}, \bibinfo {author} {\bibfnamefont {T.~S.}\ \bibnamefont {Khaire}}, \bibinfo {author} {\bibfnamefont {L.}~\bibnamefont {Knox}}, \bibinfo {author} {\bibfnamefont {A.~M.}\ \bibnamefont {Kofman}}, \bibinfo {author} {\bibfnamefont {M.}~\bibnamefont {Korman}}, \bibinfo {author} {\bibfnamefont {D.~L.}\ \bibnamefont {Kubik}}, \bibinfo {author} {\bibfnamefont {S.}~\bibnamefont {Kuhlmann}}, \bibinfo {author} {\bibfnamefont {C.-L.}\ \bibnamefont {Kuo}}, \bibinfo {author} {\bibfnamefont {A.~T.}\ \bibnamefont {Lee}}, \bibinfo {author} {\bibfnamefont {E.~M.}\ \bibnamefont {Leitch}}, \bibinfo {author} {\bibfnamefont {A.~E.}\ \bibnamefont {Lowitz}}, \bibinfo {author} {\bibfnamefont {C.}~\bibnamefont {Lu}}, \bibinfo {author} {\bibfnamefont {S.~S.}\ \bibnamefont {Meyer}}, \bibinfo {author} {\bibfnamefont {D.}~\bibnamefont
  {Michalik}}, \bibinfo {author} {\bibfnamefont {M.}~\bibnamefont {Millea}}, \bibinfo {author} {\bibfnamefont {A.}~\bibnamefont {Nadolski}}, \bibinfo {author} {\bibfnamefont {T.}~\bibnamefont {Natoli}}, \bibinfo {author} {\bibfnamefont {H.}~\bibnamefont {Nguyen}}, \bibinfo {author} {\bibfnamefont {G.~I.}\ \bibnamefont {Noble}}, \bibinfo {author} {\bibfnamefont {V.}~\bibnamefont {Novosad}}, \bibinfo {author} {\bibfnamefont {Y.}~\bibnamefont {Omori}}, \bibinfo {author} {\bibfnamefont {S.}~\bibnamefont {Padin}}, \bibinfo {author} {\bibfnamefont {Z.}~\bibnamefont {Pan}}, \bibinfo {author} {\bibfnamefont {P.}~\bibnamefont {Paschos}}, \bibinfo {author} {\bibfnamefont {J.}~\bibnamefont {Pearson}}, \bibinfo {author} {\bibfnamefont {C.~M.}\ \bibnamefont {Posada}}, \bibinfo {author} {\bibfnamefont {K.}~\bibnamefont {Prabhu}}, \bibinfo {author} {\bibfnamefont {W.}~\bibnamefont {Quan}}, \bibinfo {author} {\bibfnamefont {A.}~\bibnamefont {Rahlin}}, \bibinfo {author} {\bibfnamefont {C.~L.}\ \bibnamefont {Reichardt}},
  \bibinfo {author} {\bibfnamefont {D.}~\bibnamefont {Riebel}}, \bibinfo {author} {\bibfnamefont {B.}~\bibnamefont {Riedel}}, \bibinfo {author} {\bibfnamefont {M.}~\bibnamefont {Rouble}}, \bibinfo {author} {\bibfnamefont {J.~E.}\ \bibnamefont {Ruhl}}, \bibinfo {author} {\bibfnamefont {J.~T.}\ \bibnamefont {Sayre}}, \bibinfo {author} {\bibfnamefont {E.}~\bibnamefont {Schiappucci}}, \bibinfo {author} {\bibfnamefont {E.}~\bibnamefont {Shirokoff}}, \bibinfo {author} {\bibfnamefont {G.}~\bibnamefont {Smecher}}, \bibinfo {author} {\bibfnamefont {J.~A.}\ \bibnamefont {Sobrin}}, \bibinfo {author} {\bibfnamefont {A.~A.}\ \bibnamefont {Stark}}, \bibinfo {author} {\bibfnamefont {J.}~\bibnamefont {Stephen}}, \bibinfo {author} {\bibfnamefont {K.~T.}\ \bibnamefont {Story}}, \bibinfo {author} {\bibfnamefont {A.}~\bibnamefont {Suzuki}}, \bibinfo {author} {\bibfnamefont {K.~L.}\ \bibnamefont {Thompson}}, \bibinfo {author} {\bibfnamefont {B.}~\bibnamefont {Thorne}}, \bibinfo {author} {\bibfnamefont {C.}~\bibnamefont {Tucker}},
  \bibinfo {author} {\bibfnamefont {C.}~\bibnamefont {Umilta}}, \bibinfo {author} {\bibfnamefont {L.~R.}\ \bibnamefont {Vale}}, \bibinfo {author} {\bibfnamefont {K.}~\bibnamefont {Vanderlinde}}, \bibinfo {author} {\bibfnamefont {J.~D.}\ \bibnamefont {Vieira}}, \bibinfo {author} {\bibfnamefont {G.}~\bibnamefont {Wang}}, \bibinfo {author} {\bibfnamefont {N.}~\bibnamefont {Whitehorn}}, \bibinfo {author} {\bibfnamefont {W.~L.~K.}\ \bibnamefont {Wu}}, \bibinfo {author} {\bibfnamefont {V.}~\bibnamefont {Yefremenko}}, \bibinfo {author} {\bibfnamefont {K.~W.}\ \bibnamefont {Yoon}},\ and\ \bibinfo {author} {\bibfnamefont {M.~R.}\ \bibnamefont {Young}},\ }\bibfield  {title} {\bibinfo {title} {Performance and characterization of the {SPT}-{3G} digital frequency-domain multiplexed readout system using an improved noise and crosstalk model},\ }\bibfield  {journal} {\bibinfo  {journal} {Journal of Astronomical Telescopes, Instruments, and Systems}\ }\textbf {\bibinfo {volume} {8}},\ \href
  {https://doi.org/10.1117/1.JATIS.8.1.014001} {10.1117/1.JATIS.8.1.014001} (\bibinfo {year} {2022}),\ \bibinfo {note} {arXiv:2103.16017 [astro-ph]}\BibitemShut {NoStop}%
\bibitem [{\citenamefont {Barron}\ \emph {et~al.}(2021)\citenamefont {Barron}, \citenamefont {Mitchell}, \citenamefont {Groh}, \citenamefont {Arnold}, \citenamefont {Elleflot}, \citenamefont {Howe}, \citenamefont {Ito}, \citenamefont {Lee}, \citenamefont {Lowry}, \citenamefont {Anderson}, \citenamefont {Avva}, \citenamefont {Adkins}, \citenamefont {Baccigalupi}, \citenamefont {Cheung}, \citenamefont {Chinone}, \citenamefont {Jeong}, \citenamefont {Katayama}, \citenamefont {Keating}, \citenamefont {Montgomery}, \citenamefont {Nishino}, \citenamefont {Raum}, \citenamefont {Siritanasak}, \citenamefont {Suzuki}, \citenamefont {Takatori}, \citenamefont {Tsai}, \citenamefont {Westbrook},\ and\ \citenamefont {Zhou}}]{barron_integrated_2021}%
  \BibitemOpen
  \bibfield  {author} {\bibinfo {author} {\bibfnamefont {D.}~\bibnamefont {Barron}}, \bibinfo {author} {\bibfnamefont {K.}~\bibnamefont {Mitchell}}, \bibinfo {author} {\bibfnamefont {J.}~\bibnamefont {Groh}}, \bibinfo {author} {\bibfnamefont {K.}~\bibnamefont {Arnold}}, \bibinfo {author} {\bibfnamefont {T.}~\bibnamefont {Elleflot}}, \bibinfo {author} {\bibfnamefont {L.}~\bibnamefont {Howe}}, \bibinfo {author} {\bibfnamefont {J.}~\bibnamefont {Ito}}, \bibinfo {author} {\bibfnamefont {A.~T.}\ \bibnamefont {Lee}}, \bibinfo {author} {\bibfnamefont {L.~N.}\ \bibnamefont {Lowry}}, \bibinfo {author} {\bibfnamefont {A.}~\bibnamefont {Anderson}}, \bibinfo {author} {\bibfnamefont {J.}~\bibnamefont {Avva}}, \bibinfo {author} {\bibfnamefont {T.}~\bibnamefont {Adkins}}, \bibinfo {author} {\bibfnamefont {C.}~\bibnamefont {Baccigalupi}}, \bibinfo {author} {\bibfnamefont {K.}~\bibnamefont {Cheung}}, \bibinfo {author} {\bibfnamefont {Y.}~\bibnamefont {Chinone}}, \bibinfo {author} {\bibfnamefont {O.}~\bibnamefont {Jeong}},
  \bibinfo {author} {\bibfnamefont {N.}~\bibnamefont {Katayama}}, \bibinfo {author} {\bibfnamefont {B.}~\bibnamefont {Keating}}, \bibinfo {author} {\bibfnamefont {J.}~\bibnamefont {Montgomery}}, \bibinfo {author} {\bibfnamefont {H.}~\bibnamefont {Nishino}}, \bibinfo {author} {\bibfnamefont {C.}~\bibnamefont {Raum}}, \bibinfo {author} {\bibfnamefont {P.}~\bibnamefont {Siritanasak}}, \bibinfo {author} {\bibfnamefont {A.}~\bibnamefont {Suzuki}}, \bibinfo {author} {\bibfnamefont {S.}~\bibnamefont {Takatori}}, \bibinfo {author} {\bibfnamefont {C.}~\bibnamefont {Tsai}}, \bibinfo {author} {\bibfnamefont {B.}~\bibnamefont {Westbrook}},\ and\ \bibinfo {author} {\bibfnamefont {Y.}~\bibnamefont {Zhou}},\ }\bibfield  {title} {\bibinfo {title} {Integrated {Electrical} {Properties} of the {Frequency} {Multiplexed} {Cryogenic} {Readout} {System} for {Polarbear}/{Simons} {Array}},\ }\href {https://doi.org/10.1109/TASC.2021.3067190} {\bibfield  {journal} {\bibinfo  {journal} {IEEE Transactions on Applied Superconductivity}\
  }\textbf {\bibinfo {volume} {31}},\ \bibinfo {pages} {1} (\bibinfo {year} {2021})},\ \bibinfo {note} {conference Name: IEEE Transactions on Applied Superconductivity}\BibitemShut {NoStop}%
\bibitem [{\citenamefont {Farias}\ \emph {et~al.}(2022)\citenamefont {Farias}, \citenamefont {Russell}, \citenamefont {Kaneko}, \citenamefont {Takatori}, \citenamefont {Lee}, \citenamefont {Arnold}, \citenamefont {Adkins}, \citenamefont {Barron}, \citenamefont {Crowley}, \citenamefont {Elleflot}, \citenamefont {Fujino}, \citenamefont {Hasegawa}, \citenamefont {Ito}, \citenamefont {Lowry}, \citenamefont {Nishinomiya}, \citenamefont {Raum}, \citenamefont {Siritanasak}, \citenamefont {Westbrook},\ and\ \citenamefont {Yamada}}]{farias_-site_2022}%
  \BibitemOpen
  \bibfield  {author} {\bibinfo {author} {\bibfnamefont {N.}~\bibnamefont {Farias}}, \bibinfo {author} {\bibfnamefont {M.}~\bibnamefont {Russell}}, \bibinfo {author} {\bibfnamefont {D.}~\bibnamefont {Kaneko}}, \bibinfo {author} {\bibfnamefont {S.}~\bibnamefont {Takatori}}, \bibinfo {author} {\bibfnamefont {A.~T.}\ \bibnamefont {Lee}}, \bibinfo {author} {\bibfnamefont {K.}~\bibnamefont {Arnold}}, \bibinfo {author} {\bibfnamefont {T.}~\bibnamefont {Adkins}}, \bibinfo {author} {\bibfnamefont {D.~R.}\ \bibnamefont {Barron}}, \bibinfo {author} {\bibfnamefont {K.~T.}\ \bibnamefont {Crowley}}, \bibinfo {author} {\bibfnamefont {T.}~\bibnamefont {Elleflot}}, \bibinfo {author} {\bibfnamefont {T.}~\bibnamefont {Fujino}}, \bibinfo {author} {\bibfnamefont {M.}~\bibnamefont {Hasegawa}}, \bibinfo {author} {\bibfnamefont {J.}~\bibnamefont {Ito}}, \bibinfo {author} {\bibfnamefont {L.~N.}\ \bibnamefont {Lowry}}, \bibinfo {author} {\bibfnamefont {Y.}~\bibnamefont {Nishinomiya}}, \bibinfo {author} {\bibfnamefont
  {C.}~\bibnamefont {Raum}}, \bibinfo {author} {\bibfnamefont {P.}~\bibnamefont {Siritanasak}}, \bibinfo {author} {\bibfnamefont {B.}~\bibnamefont {Westbrook}},\ and\ \bibinfo {author} {\bibfnamefont {K.}~\bibnamefont {Yamada}},\ }\bibfield  {title} {\bibinfo {title} {On-site detector noise characterization of the {POLARBEAR}-{2A} receiver: {Millimeter}, {Submillimeter}, and {Far}-{Infrared} {Detectors} and {Instrumentation} for {Astronomy} {XI} 2022},\ }\bibfield  {journal} {\bibinfo  {journal} {Millimeter, Submillimeter, and Far-Infrared Detectors and Instrumentation for Astronomy XI}\ }\bibinfo {series} {Proceedings of {SPIE} - {The} {International} {Society} for {Optical} {Engineering}},\ \href {https://doi.org/10.1117/12.2627513} {10.1117/12.2627513} (\bibinfo {year} {2022}),\ \bibinfo {note} {publisher: SPIE}\BibitemShut {NoStop}%
\bibitem [{\citenamefont {Ghigna}\ \emph {et~al.}(2024)\citenamefont {Ghigna}, \citenamefont {Adler}, \citenamefont {Aizawa}, \citenamefont {Akamatsu}, \citenamefont {Akizawa}, \citenamefont {Allys}, \citenamefont {Anand}, \citenamefont {Aumont}, \citenamefont {Austermann}, \citenamefont {Azzoni}, \citenamefont {Baccigalupi}, \citenamefont {Ballardini}, \citenamefont {Banday}, \citenamefont {Barreiro}, \citenamefont {Bartolo}, \citenamefont {Basak}, \citenamefont {Basyrov}, \citenamefont {Beckman}, \citenamefont {Bersanelli}, \citenamefont {Bortolami}, \citenamefont {Bouchet}, \citenamefont {Brinckmann}, \citenamefont {Campeti}, \citenamefont {Carinos}, \citenamefont {Carones}, \citenamefont {Casas}, \citenamefont {Cheung}, \citenamefont {Chinone}, \citenamefont {Clermont}, \citenamefont {Columbro}, \citenamefont {Coppolecchia}, \citenamefont {Curtis}, \citenamefont {de~Bernardis}, \citenamefont {de~Haan}, \citenamefont {de~la Hoz}, \citenamefont {De~Petris}, \citenamefont {Della~Torre}, \citenamefont
  {Delle~Monache}, \citenamefont {Di~Giorgi}, \citenamefont {Dickinson}, \citenamefont {Diego-Palazuelos}, \citenamefont {Díaz~García}, \citenamefont {Dobbs}, \citenamefont {Dotani}, \citenamefont {D'Alessandro}, \citenamefont {Eriksen}, \citenamefont {Errard}, \citenamefont {Essinger-Hileman}, \citenamefont {Farias}, \citenamefont {Ferreira}, \citenamefont {Franceschet}, \citenamefont {Fuskeland}, \citenamefont {Galloni}, \citenamefont {Galloway}, \citenamefont {Ganga}, \citenamefont {Gerbino}, \citenamefont {Gervasi}, \citenamefont {Génova-Santos}, \citenamefont {Giardiello}, \citenamefont {Gimeno-Amo}, \citenamefont {Gjerløw}, \citenamefont {González~González}, \citenamefont {Grandsire}, \citenamefont {Gruppuso}, \citenamefont {Halverson}, \citenamefont {Hargrave}, \citenamefont {Harper}, \citenamefont {Hazumi}, \citenamefont {Henrot-Versillé}, \citenamefont {Hergt}, \citenamefont {Herranz}, \citenamefont {Hivon}, \citenamefont {Hlozek}, \citenamefont {Hoang}, \citenamefont {Hubmayr}, \citenamefont
  {Ichiki}, \citenamefont {Ikuma}, \citenamefont {Ishino}, \citenamefont {Jaehnig}, \citenamefont {Jost}, \citenamefont {Kohri}, \citenamefont {Konishi}, \citenamefont {Lamagna}, \citenamefont {Lattanzi}, \citenamefont {Leloup}, \citenamefont {Levrier}, \citenamefont {Lonappan}, \citenamefont {Luzzi}, \citenamefont {Macias-Perez}, \citenamefont {Maffei}, \citenamefont {Marchitelli}, \citenamefont {Martínez-González}, \citenamefont {Masi}, \citenamefont {Matarrese}, \citenamefont {Matsumura}, \citenamefont {Micheli}, \citenamefont {Migliaccio}, \citenamefont {Monelli}, \citenamefont {Montier}, \citenamefont {Morgante}, \citenamefont {Mousset}, \citenamefont {Nagano}, \citenamefont {Nagata}, \citenamefont {Natoli}, \citenamefont {Novelli}, \citenamefont {Noviello}, \citenamefont {Obata}, \citenamefont {Occhiuzzi}, \citenamefont {Odagiri}, \citenamefont {Omae}, \citenamefont {Pagano}, \citenamefont {Paiella}, \citenamefont {Paoletti}, \citenamefont {Pascual-Cisneros}, \citenamefont {Patanchon}, \citenamefont
  {Pavlidou}, \citenamefont {Piacentini}, \citenamefont {Piat}, \citenamefont {Piccirilli}, \citenamefont {Pinchera}, \citenamefont {Pisano}, \citenamefont {Porcelli}, \citenamefont {Raffuzzi}, \citenamefont {Raum}, \citenamefont {Remazeilles}, \citenamefont {Ritacco}, \citenamefont {Rubino-Martin}, \citenamefont {Ruiz-Granda}, \citenamefont {Sakurai}, \citenamefont {Savini}, \citenamefont {Scott}, \citenamefont {Sekimoto}, \citenamefont {Shiraishi}, \citenamefont {Signorelli}, \citenamefont {Stever}, \citenamefont {Sullivan}, \citenamefont {Suzuki}, \citenamefont {Takaku}, \citenamefont {Takakura}, \citenamefont {Takakura}, \citenamefont {Tartari}, \citenamefont {Tassis}, \citenamefont {Thompson}, \citenamefont {Tomasi}, \citenamefont {Tristram}, \citenamefont {Tucker}, \citenamefont {Vacher}, \citenamefont {van Tent}, \citenamefont {Vielva}, \citenamefont {Watanuki}, \citenamefont {Wehus}, \citenamefont {Westbrook}, \citenamefont {Weymann-Despres}, \citenamefont {Winter}, \citenamefont {Wollack},
  \citenamefont {Zacchei}, \citenamefont {Zannoni}, \citenamefont {Zhou},\ and\ \citenamefont {{the LiteBIRD Collaboration}}}]{ghigna_litebird_2024}%
  \BibitemOpen
  \bibfield  {author} {\bibinfo {author} {\bibfnamefont {T.}~\bibnamefont {Ghigna}}, \bibinfo {author} {\bibfnamefont {A.}~\bibnamefont {Adler}}, \bibinfo {author} {\bibfnamefont {K.}~\bibnamefont {Aizawa}}, \bibinfo {author} {\bibfnamefont {H.}~\bibnamefont {Akamatsu}}, \bibinfo {author} {\bibfnamefont {R.}~\bibnamefont {Akizawa}}, \bibinfo {author} {\bibfnamefont {E.}~\bibnamefont {Allys}}, \bibinfo {author} {\bibfnamefont {A.}~\bibnamefont {Anand}}, \bibinfo {author} {\bibfnamefont {J.}~\bibnamefont {Aumont}}, \bibinfo {author} {\bibfnamefont {J.}~\bibnamefont {Austermann}}, \bibinfo {author} {\bibfnamefont {S.}~\bibnamefont {Azzoni}}, \bibinfo {author} {\bibfnamefont {C.}~\bibnamefont {Baccigalupi}}, \bibinfo {author} {\bibfnamefont {M.}~\bibnamefont {Ballardini}}, \bibinfo {author} {\bibfnamefont {A.~J.}\ \bibnamefont {Banday}}, \bibinfo {author} {\bibfnamefont {R.~B.}\ \bibnamefont {Barreiro}}, \bibinfo {author} {\bibfnamefont {N.}~\bibnamefont {Bartolo}}, \bibinfo {author} {\bibfnamefont
  {S.}~\bibnamefont {Basak}}, \bibinfo {author} {\bibfnamefont {A.}~\bibnamefont {Basyrov}}, \bibinfo {author} {\bibfnamefont {S.}~\bibnamefont {Beckman}}, \bibinfo {author} {\bibfnamefont {M.}~\bibnamefont {Bersanelli}}, \bibinfo {author} {\bibfnamefont {M.}~\bibnamefont {Bortolami}}, \bibinfo {author} {\bibfnamefont {F.}~\bibnamefont {Bouchet}}, \bibinfo {author} {\bibfnamefont {T.}~\bibnamefont {Brinckmann}}, \bibinfo {author} {\bibfnamefont {P.}~\bibnamefont {Campeti}}, \bibinfo {author} {\bibfnamefont {E.}~\bibnamefont {Carinos}}, \bibinfo {author} {\bibfnamefont {A.}~\bibnamefont {Carones}}, \bibinfo {author} {\bibfnamefont {F.~J.}\ \bibnamefont {Casas}}, \bibinfo {author} {\bibfnamefont {K.}~\bibnamefont {Cheung}}, \bibinfo {author} {\bibfnamefont {Y.}~\bibnamefont {Chinone}}, \bibinfo {author} {\bibfnamefont {L.}~\bibnamefont {Clermont}}, \bibinfo {author} {\bibfnamefont {F.}~\bibnamefont {Columbro}}, \bibinfo {author} {\bibfnamefont {A.}~\bibnamefont {Coppolecchia}}, \bibinfo {author} {\bibfnamefont
  {D.}~\bibnamefont {Curtis}}, \bibinfo {author} {\bibfnamefont {P.}~\bibnamefont {de~Bernardis}}, \bibinfo {author} {\bibfnamefont {T.}~\bibnamefont {de~Haan}}, \bibinfo {author} {\bibfnamefont {E.}~\bibnamefont {de~la Hoz}}, \bibinfo {author} {\bibfnamefont {M.}~\bibnamefont {De~Petris}}, \bibinfo {author} {\bibfnamefont {S.}~\bibnamefont {Della~Torre}}, \bibinfo {author} {\bibfnamefont {G.}~\bibnamefont {Delle~Monache}}, \bibinfo {author} {\bibfnamefont {E.}~\bibnamefont {Di~Giorgi}}, \bibinfo {author} {\bibfnamefont {C.}~\bibnamefont {Dickinson}}, \bibinfo {author} {\bibfnamefont {P.}~\bibnamefont {Diego-Palazuelos}}, \bibinfo {author} {\bibfnamefont {J.~J.}\ \bibnamefont {Díaz~García}}, \bibinfo {author} {\bibfnamefont {M.}~\bibnamefont {Dobbs}}, \bibinfo {author} {\bibfnamefont {T.}~\bibnamefont {Dotani}}, \bibinfo {author} {\bibfnamefont {G.}~\bibnamefont {D'Alessandro}}, \bibinfo {author} {\bibfnamefont {H.~K.}\ \bibnamefont {Eriksen}}, \bibinfo {author} {\bibfnamefont {J.}~\bibnamefont {Errard}},
  \bibinfo {author} {\bibfnamefont {T.}~\bibnamefont {Essinger-Hileman}}, \bibinfo {author} {\bibfnamefont {N.}~\bibnamefont {Farias}}, \bibinfo {author} {\bibfnamefont {E.}~\bibnamefont {Ferreira}}, \bibinfo {author} {\bibfnamefont {C.}~\bibnamefont {Franceschet}}, \bibinfo {author} {\bibfnamefont {U.}~\bibnamefont {Fuskeland}}, \bibinfo {author} {\bibfnamefont {G.}~\bibnamefont {Galloni}}, \bibinfo {author} {\bibfnamefont {M.}~\bibnamefont {Galloway}}, \bibinfo {author} {\bibfnamefont {K.}~\bibnamefont {Ganga}}, \bibinfo {author} {\bibfnamefont {M.}~\bibnamefont {Gerbino}}, \bibinfo {author} {\bibfnamefont {M.}~\bibnamefont {Gervasi}}, \bibinfo {author} {\bibfnamefont {R.~T.}\ \bibnamefont {Génova-Santos}}, \bibinfo {author} {\bibfnamefont {S.}~\bibnamefont {Giardiello}}, \bibinfo {author} {\bibfnamefont {C.}~\bibnamefont {Gimeno-Amo}}, \bibinfo {author} {\bibfnamefont {E.}~\bibnamefont {Gjerløw}}, \bibinfo {author} {\bibfnamefont {R.}~\bibnamefont {González~González}}, \bibinfo {author} {\bibfnamefont
  {L.}~\bibnamefont {Grandsire}}, \bibinfo {author} {\bibfnamefont {A.}~\bibnamefont {Gruppuso}}, \bibinfo {author} {\bibfnamefont {N.~W.}\ \bibnamefont {Halverson}}, \bibinfo {author} {\bibfnamefont {P.}~\bibnamefont {Hargrave}}, \bibinfo {author} {\bibfnamefont {S.~E.}\ \bibnamefont {Harper}}, \bibinfo {author} {\bibfnamefont {M.}~\bibnamefont {Hazumi}}, \bibinfo {author} {\bibfnamefont {S.}~\bibnamefont {Henrot-Versillé}}, \bibinfo {author} {\bibfnamefont {L.~T.}\ \bibnamefont {Hergt}}, \bibinfo {author} {\bibfnamefont {D.}~\bibnamefont {Herranz}}, \bibinfo {author} {\bibfnamefont {E.}~\bibnamefont {Hivon}}, \bibinfo {author} {\bibfnamefont {R.~A.}\ \bibnamefont {Hlozek}}, \bibinfo {author} {\bibfnamefont {T.~D.}\ \bibnamefont {Hoang}}, \bibinfo {author} {\bibfnamefont {J.}~\bibnamefont {Hubmayr}}, \bibinfo {author} {\bibfnamefont {K.}~\bibnamefont {Ichiki}}, \bibinfo {author} {\bibfnamefont {K.}~\bibnamefont {Ikuma}}, \bibinfo {author} {\bibfnamefont {H.}~\bibnamefont {Ishino}}, \bibinfo {author}
  {\bibfnamefont {G.}~\bibnamefont {Jaehnig}}, \bibinfo {author} {\bibfnamefont {B.}~\bibnamefont {Jost}}, \bibinfo {author} {\bibfnamefont {K.}~\bibnamefont {Kohri}}, \bibinfo {author} {\bibfnamefont {K.}~\bibnamefont {Konishi}}, \bibinfo {author} {\bibfnamefont {L.}~\bibnamefont {Lamagna}}, \bibinfo {author} {\bibfnamefont {M.}~\bibnamefont {Lattanzi}}, \bibinfo {author} {\bibfnamefont {C.}~\bibnamefont {Leloup}}, \bibinfo {author} {\bibfnamefont {F.}~\bibnamefont {Levrier}}, \bibinfo {author} {\bibfnamefont {A.~I.}\ \bibnamefont {Lonappan}}, \bibinfo {author} {\bibfnamefont {G.}~\bibnamefont {Luzzi}}, \bibinfo {author} {\bibfnamefont {J.}~\bibnamefont {Macias-Perez}}, \bibinfo {author} {\bibfnamefont {B.}~\bibnamefont {Maffei}}, \bibinfo {author} {\bibfnamefont {E.}~\bibnamefont {Marchitelli}}, \bibinfo {author} {\bibfnamefont {E.}~\bibnamefont {Martínez-González}}, \bibinfo {author} {\bibfnamefont {S.}~\bibnamefont {Masi}}, \bibinfo {author} {\bibfnamefont {S.}~\bibnamefont {Matarrese}}, \bibinfo
  {author} {\bibfnamefont {T.}~\bibnamefont {Matsumura}}, \bibinfo {author} {\bibfnamefont {S.}~\bibnamefont {Micheli}}, \bibinfo {author} {\bibfnamefont {M.}~\bibnamefont {Migliaccio}}, \bibinfo {author} {\bibfnamefont {M.}~\bibnamefont {Monelli}}, \bibinfo {author} {\bibfnamefont {L.}~\bibnamefont {Montier}}, \bibinfo {author} {\bibfnamefont {G.}~\bibnamefont {Morgante}}, \bibinfo {author} {\bibfnamefont {L.}~\bibnamefont {Mousset}}, \bibinfo {author} {\bibfnamefont {Y.}~\bibnamefont {Nagano}}, \bibinfo {author} {\bibfnamefont {R.}~\bibnamefont {Nagata}}, \bibinfo {author} {\bibfnamefont {P.}~\bibnamefont {Natoli}}, \bibinfo {author} {\bibfnamefont {A.}~\bibnamefont {Novelli}}, \bibinfo {author} {\bibfnamefont {F.}~\bibnamefont {Noviello}}, \bibinfo {author} {\bibfnamefont {I.}~\bibnamefont {Obata}}, \bibinfo {author} {\bibfnamefont {A.}~\bibnamefont {Occhiuzzi}}, \bibinfo {author} {\bibfnamefont {K.}~\bibnamefont {Odagiri}}, \bibinfo {author} {\bibfnamefont {R.}~\bibnamefont {Omae}}, \bibinfo {author}
  {\bibfnamefont {L.}~\bibnamefont {Pagano}}, \bibinfo {author} {\bibfnamefont {A.}~\bibnamefont {Paiella}}, \bibinfo {author} {\bibfnamefont {D.}~\bibnamefont {Paoletti}}, \bibinfo {author} {\bibfnamefont {G.}~\bibnamefont {Pascual-Cisneros}}, \bibinfo {author} {\bibfnamefont {G.}~\bibnamefont {Patanchon}}, \bibinfo {author} {\bibfnamefont {V.}~\bibnamefont {Pavlidou}}, \bibinfo {author} {\bibfnamefont {F.}~\bibnamefont {Piacentini}}, \bibinfo {author} {\bibfnamefont {M.}~\bibnamefont {Piat}}, \bibinfo {author} {\bibfnamefont {G.}~\bibnamefont {Piccirilli}}, \bibinfo {author} {\bibfnamefont {M.}~\bibnamefont {Pinchera}}, \bibinfo {author} {\bibfnamefont {G.}~\bibnamefont {Pisano}}, \bibinfo {author} {\bibfnamefont {L.}~\bibnamefont {Porcelli}}, \bibinfo {author} {\bibfnamefont {N.}~\bibnamefont {Raffuzzi}}, \bibinfo {author} {\bibfnamefont {C.}~\bibnamefont {Raum}}, \bibinfo {author} {\bibfnamefont {M.}~\bibnamefont {Remazeilles}}, \bibinfo {author} {\bibfnamefont {A.}~\bibnamefont {Ritacco}}, \bibinfo
  {author} {\bibfnamefont {J.}~\bibnamefont {Rubino-Martin}}, \bibinfo {author} {\bibfnamefont {M.}~\bibnamefont {Ruiz-Granda}}, \bibinfo {author} {\bibfnamefont {Y.}~\bibnamefont {Sakurai}}, \bibinfo {author} {\bibfnamefont {G.}~\bibnamefont {Savini}}, \bibinfo {author} {\bibfnamefont {D.}~\bibnamefont {Scott}}, \bibinfo {author} {\bibfnamefont {Y.}~\bibnamefont {Sekimoto}}, \bibinfo {author} {\bibfnamefont {M.}~\bibnamefont {Shiraishi}}, \bibinfo {author} {\bibfnamefont {G.}~\bibnamefont {Signorelli}}, \bibinfo {author} {\bibfnamefont {S.~L.}\ \bibnamefont {Stever}}, \bibinfo {author} {\bibfnamefont {R.~M.}\ \bibnamefont {Sullivan}}, \bibinfo {author} {\bibfnamefont {A.}~\bibnamefont {Suzuki}}, \bibinfo {author} {\bibfnamefont {R.}~\bibnamefont {Takaku}}, \bibinfo {author} {\bibfnamefont {H.}~\bibnamefont {Takakura}}, \bibinfo {author} {\bibfnamefont {S.}~\bibnamefont {Takakura}}, \bibinfo {author} {\bibfnamefont {Y.~T.~A.}\ \bibnamefont {Tartari}}, \bibinfo {author} {\bibfnamefont {K.}~\bibnamefont
  {Tassis}}, \bibinfo {author} {\bibfnamefont {K.~L.}\ \bibnamefont {Thompson}}, \bibinfo {author} {\bibfnamefont {M.}~\bibnamefont {Tomasi}}, \bibinfo {author} {\bibfnamefont {M.}~\bibnamefont {Tristram}}, \bibinfo {author} {\bibfnamefont {C.}~\bibnamefont {Tucker}}, \bibinfo {author} {\bibfnamefont {L.}~\bibnamefont {Vacher}}, \bibinfo {author} {\bibfnamefont {B.}~\bibnamefont {van Tent}}, \bibinfo {author} {\bibfnamefont {P.}~\bibnamefont {Vielva}}, \bibinfo {author} {\bibfnamefont {K.}~\bibnamefont {Watanuki}}, \bibinfo {author} {\bibfnamefont {I.~K.}\ \bibnamefont {Wehus}}, \bibinfo {author} {\bibfnamefont {B.}~\bibnamefont {Westbrook}}, \bibinfo {author} {\bibfnamefont {G.}~\bibnamefont {Weymann-Despres}}, \bibinfo {author} {\bibfnamefont {B.}~\bibnamefont {Winter}}, \bibinfo {author} {\bibfnamefont {E.~J.}\ \bibnamefont {Wollack}}, \bibinfo {author} {\bibfnamefont {A.}~\bibnamefont {Zacchei}}, \bibinfo {author} {\bibfnamefont {M.}~\bibnamefont {Zannoni}}, \bibinfo {author} {\bibfnamefont
  {Y.}~\bibnamefont {Zhou}},\ and\ \bibinfo {author} {\bibnamefont {{the LiteBIRD Collaboration}}},\ }\href {https://doi.org/10.48550/arXiv.2406.02724} {\bibinfo {title} {The {LiteBIRD} mission to explore cosmic inflation}} (\bibinfo {year} {2024}),\ \bibinfo {note} {publication Title: arXiv e-prints ADS Bibcode: 2024arXiv240602724G}\BibitemShut {NoStop}%
\end{thebibliography}%

\end{document}